\documentclass[12pt]{amsart}

\pdfoutput = 1

\usepackage{macros}
\usepackage[normalem]{ulem}
\usepackage{float}
\usetikzlibrary{backgrounds,calc}

\usepackage{tikz-feynman}

\begin{document}
\onehalfspacing
	
\author{Fabian Hahner, Natalie M. Paquette}

\email{fhahner@uw.edu}
\email{npaquett@uw.edu}

\address{University of Washington, Department of Physics \\ 3910 15th Ave NE, Seattle, WA 98195,  U.S.A}

\title{Twisting IKKT with branes}

\begin{abstract}
    Motivated by the conjectural relationship between the IKKT matrix model and type IIB string theory, we classify supersymmetric twists of the matrix model, allowing for both dynamical and kinematical supercharges. Beyond the two twists available at any finite $N$, which we studied in our previous work, we find five additional classes at infinite rank which require non-commuting backgrounds associated with D-branes in IIB supergravity. Together, these twists realize representatives of every equivalence class of twisting supercharges in flat-space type IIB supergravity recently classified by the first author; however, the precise form of the orbit stratification differs. This difference is related to a higher central extension of the IIB superalgebra that appears in our treatment of the IKKT model in the BV formalism. Expanding around the non-commutative backgrounds gives worldvolume theories on products of A- and B-branes, including four-dimensional BF theory on an emergent D3-brane. We relate all these twists to deformations of BCOV theory (the closed string field theory of the topological B-model) using the open-closed duality of classical string field theory. For one of the twists with Spin$(7)$ symmetry, we also show the localization of the IKKT partition function by Moore, Nekrasov, and Shatashvili can be directly reconstructed with the BV formalism, with their auxiliary field completion encoded by quadratic antifield terms. Throughout, we provide self-contained treatments of the formal techniques we employ.
\end{abstract}
	
	\maketitle
	\thispagestyle{empty}
	
	\setcounter{tocdepth}{2}
	
        \newpage
	\tableofcontents
	
	\setlength{\parskip}{7pt}

\section{Introduction}

Motivated by the IKKT proposal \cite{IKKT} and related holographic proposals \cite{Hartnoll:2024csr, Ciceri:2025maa, Ciceri:2025wpb, Komatsu:2024bop, Komatsu:2024ydh, DHoker:2026eox, Najafian:2026wld} of type IIB string theory, we study the twists of the IKKT matrix model and their relation to twists of IIB string theory and supergravity. Several proposals inform this comparison; in addition to the original proposal of \cite{IKKT} (which arose from a matrix regularization of the type IIB Green--Schwarz action), recent work develops a Euclidean D-instanton holographic dictionary \cite{Ciceri:2025maa, Ciceri:2025wpb} along the lines of familiar gauge-gravity duality. Numerical simulations have also been employed to study aspects of the IKKT proposal, including the emergence of spacetime in the Lorentzian model \cite{Nishimura:2020blu,Anagnostopoulos:2022dak}. Recently, a controlled 16-supercharge-preserving deformation of the IKKT model called the polarized IKKT model has also come to the fore; there have been extensive tests of the polarized duality \cite{Hartnoll:2024csr, Komatsu:2024bop, Komatsu:2024ydh, DHoker:2026eox}. In this work, we identify the classical twisted branes that arise from large-$N$ vacua of the twisted matrix model, and compute their worldvolume theories and the closed-string theories to which they couple. In this way, we realize the open-string theories of branes in the corresponding twisted type IIB backgrounds.
 
In the IKKT model~\cite{IKKT}, the 32 supercharges of the type IIB supersymmetry algebra decompose into two classes: 16 dynamical supercharges act on the fields as determined by the dimensional reduction of ten-dimensional super Yang--Mills theory, while 16 kinematical supercharges only act by shifting the fermionic fields by a multiple of the identity. In~\cite{tw-bfss-ikkt}, we computed the two possible twists of the IKKT model that exist at any rank $N$; equivalently, these are the twists that use only dynamical supercharges. In this work, we classify all twists of the IKKT model, allowing for both dynamical and kinematical supercharges to be non-vanishing\footnote{Note that kinematical supersymmetries are absent in the polarized IKKT model, although one can study the dynamical twists of that theory in the BV formalism described below.}. In particular, new twists appear strictly at infinite $N$.

The basic mechanism that leads to these twisting backgrounds can be understood as follows. Recall that the variation of the fermion in the IKKT model under a supercharge $Q = \ekin+ \edyn$ is:
\begin{equation}\label{eq:fermvariation}
    \delta \psi = \varepsilon_{\mathrm{kin}} + \frac{1}{2} \varepsilon_{\mathrm{dyn}} \Gamma_{IJ} [X^I, X^J].
\end{equation}
Hence, a supercharge with a non-vanishing kinematical component $\ekin$ can only be preserved if its dynamical component $\edyn$ is non-vanishing as well and the matrix background has central commutators $[X^I, X^J] \sim \id$. Such backgrounds only exist at infinite $N$ and correspond to brane configurations in the IKKT model~\cite{Aoki:1999vr}.

We classify such twisting backgrounds by starting from the decomposition of the space of twisting supercharges $Q= \edyn + \ekin$ into orbits under $\Spin(10,\C)$ and R-symmetry and then constructing the different possible brane solutions that turn these into valid twisting backgrounds of the IKKT model. Then we derive the fluctuations of the IKKT model around these twisting backgrounds and compare these to known results and conjectures about twisted supergravity.

Let us now summarize our results and the structure of the paper in more detail.

In~\S\ref{sec: classif}, we introduce the IKKT model and show how dynamical and kinematical supersymmetries can be described in the BV formalism. The BV formulation of the IKKT model (in perturbation theory around a given background) retains the fluctuation fields, their gauge transformations, and their interactions in a common cohomological framework, including the terms required when supersymmetry closes only modulo equations of motion. This allows us to 1.) derive the twisted worldvolume theories of the emergent branes in ten-dimensional spacetime, and 2.) track the closed-string states to which they couple. In this formalism, we also find that, once trace modes are included, the IIB super Poincar\'e algebra is necessarily extended by a degree three cocycle to form a super Lie 2-algebra which takes the form of a $(p,q)$-string extension with $p=iq$.

A twisting background is a critical point of the BV action where the ghosts for the supersymmetry transformations take a non-vanishing value. Varying the BV action, we find the three conditions~\eqref{eq: tw-backgr} on the twisting supercharge and the matrix background: $Q$ squares to zero in the type IIB algebra, the matrix background is invariant under $Q$, and the matrices are orthogonal to the vector bilinear of $\edyn$. We then classify solutions to these conditions up to the action of $\textrm{Spin}(10, \mathbb{C})$ and R-symmetry.

In total, we find seven orbits of twisting supercharges that can be realized in the IKKT model, summarized in Table~\ref{tab: tw-backgrounds}. We describe the orbit stratification and compare it with that of the moduli space of twists of type IIB supergravity in a flat background. Every orbit of the type IIB nilpotence variety~\cite{iib-nv} has a realization in the IKKT model, but the stratification differs from type IIB supergravity (see Figure~\ref{fig:hasse-ikkt-iib}). Specifically, the $\mathrm{Sp}(4)$-orbit splits into two different orbits.

Each twisting supercharge comes attached with multiple brane backgrounds, which we describe. Computing their worldvolumes and fluxes, we show that for a pure dynamical supercharge these branes reproduce A- and B-branes in the bulk, including coisotropic A-branes that are equipped with a holomorphic-symplectic form on their worldvolume. The results are summarized in Table~\ref{tab: brane-spectra}.

In~\S\ref{sec: twists}, we compute the fluctuations around these twisting backgrounds, describe the resulting twisted theories and compare them to known results and conjectures about twisted type IIB supergravity.  Recall \cite{tw-bfss-ikkt} that the minimal twist of the IKKT model exists at any $N$ and its BV complex is identified with a fermionic, or exterior, algebra in five anticommuting variables,
\begin{equation}
    A_N = \CC[\theta^1, \dots, \theta^5] \otimes \mathfrak{gl}_N.
\end{equation}
This can be viewed as the space of open strings on a stack of $N$ point branes in the B-model on $\CC^5$ \cite{CLBCOV1}. Starting from the minimal twist, all other twists of the IKKT model can be understood through the following types of further deformations:
\begin{itemize}
    \item[---] Twisting backgrounds with non-commutative directions along a complex subspace. Here, the noncommutative configuration of matrices describes $k$ higher dimensional B-branes placed along $\CC^r \subset \CC^5$ and the worldvolume theory is a holomorphic Chern--Simons theory with BV complex
    \begin{equation}
        \Omega^{0,\bu}(\CC^r)[\theta^{r+1}, \dots , \theta^5] \otimes \mathfrak{gl}_k.
    \end{equation}
    Note that the underlying closed string field theory remains BCOV theory on $\CC^5$.
    \item[---] Twisting backgrounds with non-commutative directions along a real subspace. 
    In such twists, the matrices describe a brane with worldvolume $\R^n \times \C^m$ with $n=2,4$ and $m=0,1,\dots, 5-n$. The emergent brane is given by a product of a Lagrangian A-brane and a B-brane and the worldvolume theory on the brane is given by
    \begin{equation}
        \Omega^\bu(\R^n) \otimes \Omega^{0,\bu}(\C^m)[\theta^{m+n+1}, \dots ,\theta^5] \otimes \mathfrak{gl}_k.
    \end{equation}
    This complex is equipped with a Dolbeault plus de Rham differential $\bar \partial + \d_{dR}$ and enjoys interactions inherited from the cubic open string field theory.
    \item[---] Twisting backgrounds with non-commutative directions along a coisotropic subspace. Here, brane configurations are products of a coisotropic A-brane with worldvolume $\R^{2s} \times \CC^t_I$ and a B-brane with worldvolume $\C^m$, where $(s,t) = (0,2)$ for the $\SU(3)$-twist and $(s,t) = (1,2)$ or $(0,4)$ for the $\mathrm{Sp}(4)_A$-twist.  The complex structure $I$ on $\C^t_I$ is misaligned with that of the bulk. The worldvolume theory has BV complex
    \begin{equation}
        \Omega^\bu(\R^{2s}) \otimes \Omega^{0,\bu}(\C^t_I) \otimes \Omega^{0,\bu}(\C^m)[\theta^{2s+t+m+1}, \dots ,\theta^5] \otimes \mathfrak{gl}_k
    \end{equation}
    with differential $\d_{dR} + \bar\partial_I + \bar \partial$. In contrast to the previous cases, the Moyal product of the matrix algebra does not trivialize in cohomology, so that the worldvolume $\C^t_I$ is non-commutative.
    \item[---] Twisting backgrounds which involve the deformation of the dynamical twisting supercharge to an impure spinor. These twists are trivial in perturbation theory. Mathematically, we say their BV complexes are acyclic; they are dual to linear superpotential deformations of the B-model spacetime theory. 
\end{itemize}

Although the twists with acyclic BV complexes are trivial in perturbation theory, they carry non-perturbative content. In~\S\ref{sec: spin7}, we show how an ``on-shell'' version of the computation of the IKKT partition function through supersymmetric localization~\cite{MNS} arises in an $\Omega$-deformed version of the $\Spin(7)$-twist.

Together with~\cite{tw-bfss-ikkt}, we view this work as laying the groundwork for a systematic exploration of the various IKKT and BFSS dualities in the framework of twisted holography. One can also repeat this analysis straightforwardly for the BFSS model~\cite{BFSS, Banks:1996nn}, where non-commutative twisting backgrounds should be related to the branes appearing in twisted eleven-dimensional supergravity and M-theory~\cite{CostelloMtheory2, RSW, CY2, MaxTwist}; we leave the details of this generalization to future work.

\addtocontents{toc}{\protect\setcounter{tocdepth}{1}}
\subsection*{Acknowledgments}
We would like to thank Surya Raghavendran and Ingmar Saberi for conversations and collaborations on related projects.
NP and FH are both supported by the DOE Early Career Research Program under award DE-SC0022924. NP is also supported by the Simons Foundation as part of the Simons Collaboration on Celestial Holography. This work was supported in part by a grant of access to OpenAI models through the ChatGPT for Academic Researchers program. Computations and derivations in this work were performed by the authors; ChatGPT 6 and Claude Opus 5.5 provided assistance with drafting, proofreading, checking formulas, and literature search.
\addtocontents{toc}{\protect\setcounter{tocdepth}{2}}

\section{The moduli space of twists for the IKKT model} \label{sec: classif}

In equation \eqref{eq:fermvariation}, we recorded the variation of the IKKT model fermions under supersymmetries. For finite matrices, it implies that
\begin{equation}
    \textrm{tr}(\delta \psi) = 0 = N \varepsilon_{kin} + {1 \over 2} \varepsilon_{dyn} \Gamma_{IJ} \textrm{tr}[X^I, X^J] = N \varepsilon_{kin}.
\end{equation}
Hence, a preserved supercharge with a nonzero kinematical component is impossible at finite $N$. In this section, we will study the possible constant twisting supercharges $Q = \edyn + \ekin$, together with the requisite background matrix configuration at infinite $N$. $Q$ must be square-zero, and its action on the bosonic matrix background must vanish\footnote{Note that this construction is distinct from the fuzzy-sphere saddles studied in polarized IKKT which arise at any finite $N$; these saddles could also be studied using the twists studied in~\cite{tw-bfss-ikkt}.}. We will first derive these conditions in the Batalin-Vilkovisky (BV) formalism, and classify all admissible twisting supercharges up to the action of R-symmetry. A summary of the results including overview tables and diagrams describing the stratification of twists can be found in~\S\ref{sec: summary}.

\subsection{The IKKT model and its symmetries}
Let us start by describing the IKKT matrix model~\cite{IKKT} in the BV formalism. We build on the discussion in~\cite{tw-bfss-ikkt} where the BV action for the IKKT model and the equivariant structure under dynamical supersymmetry transformations was described. In this context, equivariant structure means the compatible action of all global symmetries on the BV theory, which includes both the gauge transformations and the equations of motion appearing in their closure. In this work, we expand this picture to include kinematical supersymmetries, which will allow us to construct a larger class of twists.

The fields of the IKKT model include ten bosonic matrices $X \in V \otimes \mathfrak{u}(N)$, that transform in the vector representation $V \simeq \mathbb{C}^{10}$ of $\Spin(10,\C)$  together with sixteen fermions $\psi^\alpha \in S_+ \otimes \mathfrak{u}(N)$ that live in the positive chirality spin representation $S_+$. These matrix-valued fields transform by conjugation under the $U(N)$ gauge symmetry; in the path integral, one integrates over $\U(N)$ equivalence classes and must divide by the finite volume $\textrm{vol}(\U(N))$. 

The action functional is obtained via dimensional reduction of ten-dimensional super Yang--Mills theory.
\begin{equation} \label{eq: action-ikkt}
    S = -\frac{1}{g} \mathrm{tr} \left(  \frac{1}{4} [X^I , X^J] [X_I, X_J] - \frac{1}{2} \psi \Gamma^I [X_I, \psi] \right)
\end{equation}
The IKKT model realizes type IIB supersymmetry, which we will review next. Here and in the following, we work with complexified super Lie algebras as is common in twisted supergravity. In particular, complex coefficients are allowed in the supersymmetry algebra and its representations\footnote{See \cite{DHoker:2026eox} for another recent study of the complexified SUSY algebra, as well as the important role of choosing a Euclidean integration contour.}; reality conditions will be imposed separately when discussing real brane configurations in the sequel. Accordingly, all groups acting on the supercharges, as well as their orbits and stabilizers, are complex, and we write $\Spin(10,\C)$, $\mathrm{SL}(5,\C)$, $\Sp(4,\C)$, and so on. Groups written without $\C$, such as $\U(N)$, $\SU(n)$, and $\mathrm{USp}(4)$, denote compact real forms. These appear when we impose Hermiticity for brane backgrounds or when the stabilizer of a twisting supercharge acts on spacetime. We often label twists by the compact forms of their stabilizers as for the $\SU(3)$-twist etc.

The type IIB supertranslation algebra $\mathfrak{t}$ is composed of supercharges and ordinary translations. This can be expressed in Lie algebraic notation, where $\Pi$ denotes operators of fermionic, or odd, parity, as:
\begin{equation}
\ft = \Pi(S_+ \otimes W) \oplus V ,
\end{equation}
where $W = (\mathbb{C}^2, (-,-)_W)$ is the two-dimensional complex vector space labeling the two members of the type IIB supersymmetry doublet. It is equipped with a non-degenerate symmetric, complex bilinear form $ (-,-)_W$. The bracket between two odd elements is 
\begin{equation}
    \Gamma \otimes (-,-)_W : \Sym^2(S_+ \otimes W) \longrightarrow V,    
\end{equation}
where $\Gamma: \text{Sym}^2(S_+) \longrightarrow V$ is the vector invariant built from a pair of spinors; in a basis the components of this map are the Gamma matrices $\Gamma^I_{\alpha \beta}$. Here, $\text{Sym}^2$ denotes the symmetric tensor square. 

The super Poincar\'e algebra is given by the semidirect product $\fp = (\mathfrak{so}(10,\C) \oplus \mathfrak{o}(2,\C)) \ltimes \ft $. The full Lorentz and R-symmetry groups are given by $\mathrm{Spin}(10, \C) \times \mathrm{O}(2, \C)$ which acts on $\mathfrak{t}$, where $\mathrm{O}(2,\C)$ is the automorphism group of $(W, (-,-)_W)$. 

The space
\begin{equation}
    Y_{IIB} = \{ Q \in S_+ \otimes W \: | \: [Q,Q] = 0\}
\end{equation}
is called the \textit{nilpotence variety}~\cite{NV, ElliottSafronov} for the $\cN=(2,0)$ super Poincar\'e algebra. It is the solution set of the quadratic equations $[Q, Q]=0$\footnote{Throughout the text $[-,-]$ denotes the graded bracket of a super Lie algebra with symmetry $[A, B] = -(-1)^{F_A F_B} [B,A]$. In particular, the bracket is symmetric between two fermionic elements.} and forms the space of twists for type IIB supergravity on $\RR^{10}$. An orbit of the nilpotence variety consists of supercharges which are related by Lorentz and R-symmetries. The orbit stratification of such a variety describes these orbits and the limits in which one orbit may degenerate into another. The orbit stratification for the superalgebra of type IIB supergravity was recently described in~\cite{iib-nv}, see also~\cite{mod-type-ii}.

To relate this to the IKKT model, we choose a basis $(w, w^\vee)$ of the two-dimensional space $W$ by the properties $(w, w)_W = (w^\vee, w^\vee)_W = 0$ and $(w, w^\vee)_W = 1$. These basis vectors are isotropic, meaning they each have zero self-pairing. We decompose the odd part of the superalgebra as
\begin{equation}
    \Pi(S_+ \otimes W) = \Pi(S_+ \otimes \langle w \rangle) \oplus \Pi(S_+ \otimes \langle w^\vee \rangle),
\end{equation} where, e.g, $\langle w \rangle$ denotes the line spanned by $w$. 
Note that these two summands are two purely odd abelian subalgebras. These are called the dynamical and kinematical supersymmetries respectively and we split a general supercharge accordingly:
\begin{equation}
    Q = \edyn \otimes w + \ekin \otimes w^\vee .
\end{equation}
In particular, we note that the translations arise as brackets between kinematical and dynamical supercharges. In this basis, the square-zero condition conveniently becomes 
\begin{equation}\label{eq:squarezero}
    \edyn \Gamma^{I}\ekin=0.
\end{equation}
Finally, we note for later use that $\SO(2, \mathbb{C})$ rescales $w, w^{\vee}$ inversely, while the $\ZZ_2$ in $\mathrm{O}(2, \mathbb{C}) \cong \SO(2,\CC) \rtimes \ZZ_2$ exchanges them.

Recall that on the fields of the IKKT model, supersymmetry transformations only close up to both gauge transformations and the equations of motion. The BV formalism retains this information by adjoining both ghosts and antifields. As in the familiar BRST formalism, ghosts represent symmetry parameters with shifted grading, while antifields are partners of both physical fields and ghosts, but with opposite Grassmann parity. Below we will use $\left\lbrace -, - \right\rbrace$ to denote the BV antibracket, which is an odd Poisson bracket that pairs fields and ghosts with their antifields.

We combine the classical action together with its gauge symmetries and supersymmetry transformations into the BV action,
\begin{equation}
    S_{BV} = S + S_{\mathrm{gauge}} + S_{\mathrm{susy}}.
\end{equation} The action generates the transformation
\begin{equation}
    Q_{BV} = \left\lbrace S_{BV}, - \right\rbrace
\end{equation} called the BV differential, so named because it must square to zero. The classical master equation 
\begin{equation}
    \left\lbrace S_{BV}, S_{BV}\right\rbrace = 0
\end{equation} ensures this property at tree level.

In~\cite{tw-bfss-ikkt}, the BV action of the IKKT model was described, but with $S_{\mathrm{susy}}$ covering only the transformations under dynamical supersymmetries. In the following, we give the full treatment including kinematical supersymmetries and translations.

For the IKKT model, we expand the space of physical fields $(X^I , \psi_\alpha)$ by introducing the ghost $c \in \mathfrak{u}(N)$ together with antifields $(X^{*I} , \psi^*_\alpha)$ and the ghost antifield $c^*$. All fields transform in the adjoint of the gauge group so that $S_{\mathrm{gauge}}$ is:
\begin{equation}
    S_{\mathrm{gauge}} = \mathrm{tr}(\frac{1}{2} c^* [c,c] + X^{*I} [c, X_I] + \psi^* [c, \psi] ).
\end{equation}

To encode the equivariant structure under supersymmetries, which we will need for twisting, we introduce $\varepsilon_{\mathrm{dyn}}$ and $\varepsilon_{\mathrm{kin}}$ parameterizing dynamical and kinematical supersymmetry transformations. Under dynamical supersymmetry transformations $\varepsilon_{dyn}=: \varepsilon$, the BV fields transform as:
\begin{equation} \label{eq: dyn-susy trafos}
    \begin{array}{cc}
        \delta c = \varepsilon \Gamma^I \varepsilon X_I & \delta X^*_I = \varepsilon \Gamma_{IJ} [\psi^* , X^J] + \varepsilon \Gamma_I \varepsilon c^* \\[6pt]
        \delta X^I = \varepsilon \Gamma^I \psi & \delta \psi^* = \varepsilon \Gamma^I X^*_I \\[6pt]
        \delta \psi = \frac{1}{2} \varepsilon \Gamma_{IJ} [X^I,X^J] + (\varepsilon \Gamma^I \varepsilon) \Gamma_I \psi^* - 2\varepsilon (\varepsilon \psi^*) &   \delta c^* = 0
    \end{array}
\end{equation} where $\delta$ denotes the supersymmetry contribution to $Q_{BV}$.
These transformations, including their closure terms by gauge transformations and the fermionic equation of motion, are summarized in the following piece of the action:
\begin{equation}\label{eq: dyn-ikkt-susy}
    S_{\mathrm{susy}}^{\mathrm{dyn}} = \mathrm{tr}( X^{*}_I \varepsilon \Gamma^I \psi + \frac{1}{2} \psi^* \varepsilon \Gamma_{IJ} [X^I, X^J] + c^* (\varepsilon \Gamma_I \varepsilon)X^I + \frac{1}{2}(\varepsilon \Gamma_I \varepsilon) (\psi^* \Gamma^I \psi^*) - (\varepsilon \psi^*)^2 ).
\end{equation}
Recall that the ghost for an odd supersymmetry is an even, commuting field. In this action, $\varepsilon$ plays the role of a bosonic ghost for the dynamical supersymmetry transformations. Assigning a constant value to such a supersymmetry ghost and expanding the BV action around this value deforms the BV differential by the action of the corresponding supercharge and thus specifies a twist.

In addition, the kinematical supersymmetries act on the trace mode of the fermion as a constant shift
\begin{equation}
    \delta \psi = \varepsilon_{\mathrm{kin}} \: 1_N,
\end{equation}
while the ordinary translations shift the trace mode of the bosonic matrices
\begin{equation}
    \delta X^I = a^I \: 1_N .
\end{equation}
Thus, we have to add the following terms to the BV action to account for the full equivariant structure:
\begin{equation}
    S_{\mathrm{susy}}^{\mathrm{kin}} = \varepsilon_{\mathrm{kin}} \tr( \psi^*) + a^I \tr( X^*_I) + (\varepsilon_{\mathrm{dyn}} \Gamma^I \varepsilon_{\mathrm{kin}} ) a^*_I ,
\end{equation}
where, $a^I$ is the ghost for translations and $a^*_I$ is its antifield. 

However, after defining the BV action as
\begin{equation}
    S_{BV} = S + S_{\mathrm{gauge}} + S_{\mathrm{susy}}^{\mathrm{dyn}} + S_{\mathrm{susy}}^{\mathrm{kin}},
\end{equation}
one may perform a short computation that shows that this action does not satisfy the classical master equation. Instead, we find
\begin{equation}
    \{ S_{BV} , S_{BV} \} = (\varepsilon_{\mathrm{dyn}} \Gamma^I \varepsilon_{\mathrm{dyn}}) a_I \tr(c^*) .
\end{equation}
This arises because the gauge-transformation term in the square of a dynamical supersymmetry transformation has field-dependent parameter $\edyn\Gamma^I \edyn X_I$. The contribution of translations $a^I 1_N$ to $Q_{BV} X^I$ changes this parameter by $\edyn\Gamma^I \edyn a_I 1_N$, which is a central gauge parameter acting trivially by commutators on all matrix fields.
Since the right-hand side is proportional to $\tr(c^*)$, this obstruction is absent for gauge group $\SU(N)$. It arises because translations and kinematical supersymmetries act on the trace modes of the matrices, which are only present for $\U(N)$.

We can correct $S_{BV}$ so that it solves the classical master equation by introducing an additional ghost $b$ and deforming
\begin{equation}
    S_{BV} \mapsto S_{BV} + b\tr(c^*) + (\varepsilon_{\mathrm{dyn}} \Gamma^I \varepsilon_{\mathrm{dyn}}) a_I b^*.
\end{equation}
The new field $b$ is an even ghost with ghost number 2, and $b^*$ is its antifield. This modification has an instructive interpretation in terms of central extensions related to branes~\cite{twistedBranes, FSS}. These terms describe a central extension of the type IIB super Poincar\'e algebra to an $L_\infty$ algebra\footnote{An $L_{\infty}$ algebra allows for compatible multilinear brackets in addition to the ordinary binary bracket.} by a three-ary operation. Concretely, we have a bracket between two dynamical supercharges and a translation:
\begin{equation}
    \mu_3(Q , Q , v) = (\edyn \Gamma^I \edyn) v_I .
\end{equation}
This is precisely the form of a string extension of the type IIB supersymmetry algebra but restricted to the dynamical supercharges. Recall that every traceless symmetric bilinear form\footnote{Here, $W^{\vee}= \textrm{Hom}(W, \mathbb{C})$ is the dual space of $W$ equipped with the inverse of the bilinear pairing on $W$.} $\beta \in \Sym^2_0 (W^\vee)$, defines a string extension acting on supercharges $Q = \edyn \otimes w + \ekin \otimes w^\vee$ as
\begin{equation}
    \mu_3^\beta(Q,Q,v) = \left( \Gamma^I(\edyn , \edyn) \beta(w,w) + \Gamma^I(\ekin,\ekin) \beta(w^\vee, w^\vee) \right) v_I.
\end{equation}
In an orthonormal basis of $W$ with respect to $(-,-)_W$, we can set $w = 1/\sqrt{2} (1,i)$ and $w^\vee = 1/\sqrt{2} (1,-i)$. Expanding,
\begin{equation}
    \beta = p \begin{pmatrix}
        1 & 0 \\ 0 & -1
    \end{pmatrix} + q \begin{pmatrix}
        0 & 1 \\ 1 & 0
    \end{pmatrix}
\end{equation}
with $p, q \in \mathbb{C}$ matches with the usual conventions of the $(p,q)$-string extension and we find that our extension of the IKKT model corresponds to $\beta$ with $p = iq$.

For the readers' convenience, we summarize the ingredients of the BV action described in this section:
\begin{table}[!htbp]
\centering
\small
\setlength{\tabcolsep}{5pt}
\renewcommand{\arraystretch}{1.5}
\begin{tabular}{l|l|l|l}
Field & Role & Grassmann parity & Ghost number (before twisting) \\
\hline
$X^I$ & Bosonic matrix & Even & $0$ \\
$\psi$ & Fermionic matrix & Odd & $0$ \\
\hline
$c$ & Gauge ghost & Odd & $1$ \\
$\varepsilon_{\mathrm{dyn}},\,\varepsilon_{\mathrm{kin}}$ & Supersymmetry ghosts & Even & $1$ \\
$a^I$ & Translation ghost & Odd & $1$ \\
$b$ & Additional central ghost & Even & $2$ \\
\end{tabular}
\caption{Quantities appearing in the BV action of the IKKT matrix model. Antifields are not displayed explicitly but given a field with ghost numbers $g$, its antifield partner has ghost number $-g-1$; it also has opposite Grassmann parity.}
\label{tab:ghosts}
\end{table}

Finally, we note that this extension changes the global structure of the even part of the symmetries: While $\Spin(10,\C) \times \mathrm{O}(2,\C)$ acts on the type IIB supertranslation algebra, the R-symmetry group is broken to $\mathrm{SO}(2,\C)$. The missing $\mathbb{Z}_2$ corresponds to the transformation that exchanges kinematical and dynamical supercharges. Note that in the type IIB string, this transformation exchanges the D-instanton with the anti-D-instanton sector. Since this reflection changes the D-instanton sector underlying the matrix model, supercharges related by it need not define equivalent twists within a fixed IKKT matrix theory. This will be of significance below in the classification of twisting backgrounds.

\subsection{Twists of the IKKT model at infinite $N$}
We now discuss the conditions that define valid twisting backgrounds of the IKKT matrix model and introduce the relevant tools and notation for the classification of such backgrounds in~\S\ref{sec: class-tw-back}.

\subsubsection{Twisting backgrounds}
A twisting background is a critical point of the (equivariant) BV action of the matrix model, with a non-zero value assigned to the supersymmetry ghosts. This value is the twisting supercharge $Q = \edyn \otimes w + \ekin \otimes w^\vee$ and expanding the BV action around such a background adds $\delta_Q$ to the BV differential. The remaining variations of the BV action impose consistency conditions on $Q$ and the matrix background, which we derive below. We will consider configurations where only the supersymmetry ghosts and the $X^I$ take nonzero vacuum expectation values\footnote{Note that expanding the theory around a twisting background breaks the $\ZZ \times \ZZ_2$ grading of ghost degree and internal parity to a single $\Z_2$-degree. In some cases this $\Z_2$ can be lifted back to a $\ZZ$-grading (see for example~\cite{CostelloHol}).}.

We now collect all non-trivial variations of the actions containing the supersymmetry ghosts $(\varepsilon_\mathrm{dyn} , \varepsilon_\mathrm{kin})$. Together, these equations give the full conditions that need to be satisfied for a configuration to define a valid twisting background. First, varying with respect to the antifield of a translation ghost, we find
\begin{equation}
    \frac{\delta S_{BV}}{\delta a^*_I} = \varepsilon_\mathrm{dyn} \Gamma^I \varepsilon_\mathrm{kin} = 0.
\end{equation}
This is the square-zero condition in the type IIB supertranslation algebra expressed in the isotropic basis that distinguishes kinematical and dynamical supercharges. Solutions to this equation are the variety described in~\cite{iib-nv} that forms the space of twisting backgrounds for type IIB supergravity in a flat spacetime.

However, for the IKKT matrix model additional conditions are present. Evaluating the variation with respect to the antifield of the fermion, we find
\begin{equation}
    \frac{\delta S_{BV}}{\delta \psi^*} = \varepsilon_{\mathrm{kin}} + \frac{1}{2} \varepsilon_{\mathrm{dyn}} \Gamma_{IJ} [X^I, X^J],
\end{equation}
where we already have restricted ourselves to configurations where all fields except $\varepsilon_{\mathrm{kin}}$, $\varepsilon_{dyn}$ and $X^I$ are vanishing. Thus, we find that twisting backgrounds with non-vanishing $\varepsilon_{\mathrm{kin}}$ necessarily have to be accompanied by a non-vanishing value of $\varepsilon_\mathrm{dyn}$ and $[X_I, X_J] \sim 1$ with the latter only being possible at infinite $N$, as described in the introduction. 

Finally, varying with respect to the antifield of the ghost $c^*$ for the gauge transformations, we find
\begin{equation}
    (\varepsilon_\mathrm{dyn} \Gamma^I \varepsilon_\mathrm{dyn}) X_I = 0.
\end{equation}
This condition says that the value assigned to a bosonic matrix $X^I$ has to be orthogonal to the vector invariant of the dynamical twisting supercharge.

In summary, consistent twisting backgrounds of the IKKT model at infinite $N$ are joint solutions to the equations:
\begin{equation} \label{eq: tw-backgr}
    \qquad \varepsilon_{\mathrm{dyn}} \Gamma^I \varepsilon_{\mathrm{kin}} = 0 \qquad \quad \varepsilon_{\mathrm{kin}} + \frac{1}{2} \varepsilon_{\mathrm{dyn}} \Gamma_{IJ} [X^I, X^J] = 0  \qquad \quad (\varepsilon_\mathrm{dyn} \Gamma^I \varepsilon_\mathrm{dyn}) X_I = 0.
\end{equation}
These conditions play the following roles: the first condition renders $Q$ square-zero in the global supertranslation algebra; the second is the condition that the matrix background is preserved by $Q$; the third imposes the closure condition.
In addition, the equations of motion for the bosonic matrices $[X_J, [X^I, X^J]]$ also have to hold. We are interested in solutions where the commutator of the bosonic matrices is central, hence we will specialize to solutions of the form
\begin{equation}
    [X^I , X^J] = \Theta^{IJ} \id,
\end{equation}
for $\Theta$ an antisymmetric rank-2 tensor representation, i.e. $\Theta \in \wedge^2 V \cong \so(10)$. Further, we denote the vector bilinear of the dynamical twisting supercharge by $v^I = \edyn \Gamma^I \edyn$. Then the conditions on such a twisting background can be rewritten compactly as
\begin{equation} \label{eq: tw-back-2}
    \Theta(v) = 0 \qquad \quad \ekin + \Theta \cdot \edyn = 0 \qquad \quad v_I X^I = 0,
\end{equation}
where $\cdot$ denotes Clifford multiplication and $\Theta(v)$ is the action of $\Theta \in \mathfrak{so}(10)$ on $v$. 

Solutions are acted upon by $\mathrm{Spin}(10,\C) \times \mathrm{SO}(2,\C)$ so that twisting backgrounds decompose into orbits, with those solutions in the same orbit corresponding to equivalent twists. As discussed above, the R-symmetry group acting here is $\mathrm{SO}(2,\C)$ instead of $\mathrm{O}(2,\C)$ which acts on the nilpotence variety of the type IIB supertranslation algebra~$\ft$. The missing $\ZZ_2$ exchanges kinematical and dynamical supersymmetries; in type IIB string theory, this transformation exchanges the D-instanton with the anti-D-instanton. \footnote{Strictly speaking, $\SO(2,\C)$ acts on the IKKT model only in combination with a rescaling of the matrices, which rescales the coupling, and an overall rescaling of the supercharges. Since a twist only depends on $Q$ up to rescaling, this does not affect the classification.}

The twisted theory is the theory obtained by expanding the full BV theory around this background. We will study the BV complex of fluctuations around the admissible twisting backgrounds in~\S\ref{sec: twists}. The linearized BV differential squares to zero on these fluctuations, while higher-order terms in the differential will retain the interactions of the twisted theory.

\subsubsection{Holomorphic decomposition} \label{sec: class-twist}
We begin by choosing a five-dimensional complex subspace  of the complexified vector representation of $\textrm{Spin}(10)$, $L \subset V$, such that the bilinear pairing vanishes on every pair of vectors in $L$. This defines a maximal isotropic subspace. Together with its dual space, $L^{\vee}:= \textrm{Hom}(L, \mathbb{C})$, we can decompose $V = L \oplus L^{\vee}$. 

Upon making this choice, the spinor representation $S_+$ is identified with the even exterior powers of the dual space $L^\vee$. Below $\wedge^k L^{\vee}$ denotes the space of antisymmetric $k$-tensors, so that
\begin{equation} \label{eq: spinor-rep}
    S_+ = \wedge^{0} L^\vee \oplus \wedge^2 L^\vee \oplus \wedge^4 L^\vee,
\end{equation}
while the negative chirality spinor representation $S_-$ is built from the odd exterior powers. Note that this is nothing but the fermionic oscillator description of a chiral spinor: the scalar $1 \in \wedge^0 L^{\vee}$ is a vacuum annihilated by contraction with every vector in $L$; the differential forms encode spinor components while wedge multiplication is used to create excitations on top of the vacuum. The action of Clifford multiplication $V \times S_+ \longrightarrow S_-$ is given by $L$ acting via contraction $\iota$ and $L^\vee$ via the wedge product $\wedge$.

According to~\eqref{eq: spinor-rep}, both kinematical and dynamical supersymmetries decompose into zero, two, and four-form components $\varepsilon = (\varepsilon^{(0)} , \varepsilon^{(2)} , \varepsilon^{(4)})$. In the following, we will fix a basis $(e^1, \dots, e^5)$ of $L$ with dual basis $(e_1^\vee, \dots , e_5^\vee)$ of $L^\vee$, such that $e_i^{\vee}(e^j) = \delta_i^j$, and where we denote, e.g., $e^{\vee}_{12} = e_1^{\vee} \wedge e_2^{\vee}$. Representatives for twisting backgrounds will be written in this basis.

Further, we can decompose the ten bosonic matrices as
\begin{equation}
    X^I = Z^i e^\vee_i + \bar{Z}_i e^i.
\end{equation}
The exterior square of the vector representations decomposes as
\begin{equation}
    \wedge^2 V = \wedge^2 L^\vee \oplus (L^\vee \otimes L) \oplus \wedge^2 L
\end{equation}
with the summands corresponding to the commutators 
\begin{equation}
    \Theta^{ij} = [Z^i, Z^j] \qquad \Theta^i_j = [Z^i, \bar Z_j] \qquad \Theta_{ij} =[\bar Z_i , \bar Z_j].
\end{equation}
We will refer to the corresponding backgrounds of type $(2,0)$, $(1,1)$, and $(0,2)$ respectively. Note that the $\Theta^i_j$ further decomposes into a trace and a traceless part.

With $\Gamma_{IJ} = \tfrac12 [\Gamma_I, \Gamma_J]$ and $\{\Gamma_I, \Gamma_J\} = 2 \delta_{IJ}$, we have $X^I \Gamma_I = Z^i \, e^\vee_i \wedge {} + \bar Z_i \, \iota_{e_i}$, so that the action of $\Theta$ by Clifford multiplication on $S_+ = \wedge^{\mathrm{even}} L^\vee$ is given by
\begin{equation} \label{eq: theta-action}
    \Theta \cdot \varepsilon = \tfrac12 \Theta^{ij} \, e^\vee_i \wedge e^\vee_j \wedge \varepsilon + \Theta^i_j \, e^\vee_i \wedge \iota_{e_j} \varepsilon - \tfrac12 \Theta^i_i \, \varepsilon + \tfrac12 \Theta_{ij} \, \iota_{e_i} \iota_{e_j} \varepsilon .
\end{equation}

\subsubsection{Brane backgrounds and their worldvolumes} \label{ssec: worldvolumes}
Backgrounds that are purely of type $(1,1)$, i.e.\ with $\Theta^{(2,0)} = 0$ (and therefore, by Hermiticity, also $\Theta^{(0,2)} = 0$), describe branes along complex submanifolds of $\C^5$.
For example, $[X^1, X^2] = \tfrac{i}{2} \id$ with all other matrices vanishing gives
\begin{equation}
    [Z^1, \bar Z_1] = -2i [X^1, X^2] = 1 ,
\end{equation}
which describes a D1-brane spanning the complex line $\C_{z_1} \subset \C^5$.

On the other hand, backgrounds with non-vanishing $\Theta^{(2,0)}$ describe branes whose worldvolumes do not fully align with the complex structure. The structure and dimension of their worldvolume, however, are not determined by $\Theta^{(2,0)}$ alone, but also depend on $\Theta^{(1,1)}$. To see this, consider a background in the directions $z_1, z_2$ with
\begin{equation}
    [Z^1, Z^2] = [\bar Z_1, \bar Z_2] = i \vartheta \, \id , \qquad [Z^1, \bar Z_2] = [\bar Z_1, Z^2] = i \rho \, \id ,
\end{equation}
where $\vartheta, \rho \in \R$ and $[Z^1, \bar Z_1] = [Z^2, \bar Z_2] = 0$. In terms of the real coordinates, the only non-vanishing commutators are
\begin{equation} \label{eq: nc-real}
    [X^1, X^3] = \tfrac{i}{2} (\vartheta + \rho) \, \id , \qquad [X^2, X^4] = \tfrac{i}{2} (\rho - \vartheta) \, \id .
\end{equation}
For $\rho = \vartheta$, only $X^1$ and $X^3$ fail to commute, and we obtain a D1-brane along the totally real plane spanned by $X^1$ and $X^3$; in particular, the worldvolume is two-dimensional. For $\rho \neq \pm \vartheta$, however, both commutators are non-zero, so that all four of $X^1, \dots, X^4$ are non-commutative and the worldvolume fills all of $\R^4$. In general, if $\Theta^{(2,0)}$ is non-degenerate on $2s$ complex directions, the brane extends along as few as $2s$ or as many as $4s$ of the corresponding real directions.

When $\varepsilon_\mathrm{dyn} = 1$, the $(0,2)$ and traceless $(1,1)$ components annihilate the spinor. The remaining $(2,0)$ component and the trace then determine $\varepsilon_\mathrm{kin}$ through~\eqref{eq: tw-back-2}. (We emphasize that here the trace is taken over the five spacetime indices $\Theta^i_i$, not over gauge indices.) In the example above, the twist thus fixes $\vartheta$, while $\rho$ is free, so that the same twist admits both the totally real D1-brane and the brane which spans $\R^4$.

\subsubsection{Twisting supercharges and pure spinors}
We now introduce some terminology that is useful in the classification of twists. For a nonzero spinor $\varepsilon$, define its annihilator by
\begin{equation}
  L_{\varepsilon}:= \left\lbrace u \in V | u^I \Gamma_I \varepsilon = 0 \right\rbrace.
\end{equation}For example, $L_1 = L$.  A spinor is called \textit{pure} when $L_{\epsilon}$ has dimension 5 or, equivalently, $\Gamma(\epsilon, \epsilon) = 0$. Any nonzero spinor that is not pure is called \textit{impure}. 

Notice that purity is a condition of any individual spinor, whereas the IIB square-zero condition \eqref{eq:squarezero} concerns a pair of dynamical and kinematical spinors. A dynamical supercharge is therefore square-zero in the IIB algebra even when the dynamical spinor is impure. 

Under $\textrm{Spin}(10, \mathbb{C})$, spinors form two orbits~\cite{Igusa}. One is a smaller pure spinor orbit, and the other is a generic orbit of impure spinors. The stabilizer of a spinor is the subgroup that leaves it fixed. 
\begin{itemize}
    \item [---] The minimal orbit has dimension 11. The stabilizer of such supercharges is $\mathrm{SL}(5, \C) \ltimes \wedge^2 L$.
    \item[---] The open orbit of impure spinors has dimension 16. The stabilizer of supercharges in this orbit is $\mathrm{Spin}(7, \C) \ltimes \CC^8$.
\end{itemize}

Given a supercharge $Q$, the adjoint action $[Q, -]$ maps a second supercharge to a translation. The image of this map tells us the set of translations that are $Q$-exact, so that such translations are trivial in $Q$-cohomology. Thus $Q$ determines a subspace
\begin{equation}\label{eq:image}
    \im[Q,-] = \im(\Gamma(\ekin,-)) + \im(\Gamma(\edyn,-)) \subseteq V.
\end{equation}
The image here is taken in the ten-dimensional type IIB
supertranslation algebra. Although the IKKT model has a
zero-dimensional worldvolume, it retains these translations as shifts $\delta_a X^I=a^I\mathbf{1}_N$ of the D-instanton position variables. In the corresponding bulk theory, the same generators
act as spacetime derivatives, so their $Q$-exactness determines the topological and holomorphic directions of the bulk. Therefore, this subspace fixes the spacetime dependence of the corresponding twist of type IIB supergravity. We use the notation $\mathbb{R}^{10-2k} \times \mathbb{C}^k$ to denote a twist that is topological along the real factor and holomorphic along the complex factor. The exact translations comprise the directions of the topological factor together with the antiholomorphic translations, i.e. $\textrm{dim}_{\mathbb{C}}\textrm{im}[Q, -] = 10-k$. A particular matrix background will specify a brane worldvolume within this ambient spacetime, as we delineate below.  Note that by \eqref{eq: tw-back-2}, the kinematical supercharge is completely determined by the dynamical supercharge and the brane background together. In particular, we can rewrite \eqref{eq:image} as
\begin{equation}
    \im[Q,-] = \im(\Gamma(\edyn,-)) + \Theta(\im(\Gamma(\edyn,-))).
\end{equation}

For those readers for whom the exterior algebra notation we use is unfamiliar, let us do a small example to illustrate its utility. Consider a pure spinor $\edyn$ and use the $\Spin(10,\C)$-action to rotate it to 1. The map $\Gamma(1, -)$ vanishes on the zero and two-form components, and is injective on the four-form component. The square-zero condition then forces $\ekin = a + B$ for $a \in \mathbb{C}$ and $B \in \wedge^2 L^{\vee}$. Explicitly, $B = {1 \over 2}B_{ij}e_i^{\vee} \wedge e_j^{\vee}$ is an antisymmetric $5 \times 5$ matrix, whose rank may be 0, 2, or 4. The transformations that preserve our rotated spinor $1$ include contractions:
\begin{equation}
    a + B \mapsto a + \iota_{\eta}B + B,  \qquad \eta \in \wedge^2 L
\end{equation} where $\iota_\eta B$ contracts both indices of $B$ and produces a scalar. In particular, it can be chosen to cancel $a$ as long as $B \neq 0$. The remaining $SL(5, \mathbb{C})$ freedom can be used to bring $B$ into a rank-two or rank-four normal form. (If $B = 0$, the scalar is either zero or can be normalized to one using the Spin and R-symmetry).

\subsection{Classification of twisting backgrounds} \label{sec: class-tw-back}
In this subsection, we employ the tools established in the previous subsection to classify twisting backgrounds. We proceed by starting from the classification of twisting supercharges $(\edyn, \ekin)$ and then solving for the constant-commutator brane backgrounds that turn this configuration into a valid twisting background of the IKKT matrix model. For a given twisting supercharge, these backgrounds come in continuous families, labeled by the fluxes and by the orientation of the brane relative to the topological and holomorphic directions selected by the twisting supercharge. We describe representatives that are block diagonal with respect to these directions. For each possible worldvolume dimension a block diagonal representative exists and as we will see in~\S\ref{sec:furthertwists}, they realize all worldvolume theories. The results are summarized in~\S\ref{sec: summary}.

\subsubsection{The minimal twist and the $\Spin(7)$-twist}
First, if the kinematical part of the twisting supercharge vanishes, we recover the two twists discussed in~\cite{tw-bfss-ikkt}. The dynamical twisting supercharge can be pure or impure, and we discuss the two cases in turn. In both cases, the trivial background $X^I=0$ is allowed, so these twists exist at any $N$. Both twists also admit nonzero commuting backgrounds. Within the ansatz above, these satisfy
\[
  [X^I,X^J]=0,
  \qquad
  (\varepsilon_{\mathrm{dyn}}\Gamma^I
   \varepsilon_{\mathrm{dyn}})X_I=0.
\]
The second condition is automatic in the pure case, while in the impure case it sets $Z^5=0$ for the representative $\edyn = 1 + e^\vee_{1234}$ chosen below (and hence also $\bar Z_5=0$ when Hermiticity is imposed). As usual, at finite $N$, simultaneous diagonalization identifies commuting Hermitian backgrounds with D-instantons at the spacetime positions specified by their joint eigenvalues, and $X^I=0$ describes the coincident stack at the origin. These backgrounds do not introduce additional orbits of twisting supercharges, although a particular configuration can break part of the stabilizer of the twisting supercharge.

The minimal orbit is the pure spinor orbit, and we can take $\edyn = 1$ as a representative. The stabilizer of the trivial twisting background is $\mathrm{SL}(5, \C) \ltimes \wedge^2 L \times \C^\times \subset \Spin(10,\C) \times \SO(2,\C)$. The image of this twist $\operatorname{im}[Q,-]=L\subset V$ is a maximal isotropic five-plane. On the Euclidean real slice, identifying $L$ with the antiholomorphic translation directions selects a complex structure on the ambient $\mathbb{R}^{10}\cong\mathbb{C}^5$.
The corresponding bulk twist is therefore holomorphic; it is the minimal twist in the sense that all other twists can be obtained as further deformations from it. In keeping with the terminology of~\cite{iib-nv} we also refer to it as the holomorphic isotropic twist.

For the impure case, take $\ekin = 0$ and $\edyn = 1 + e^{\vee}_{1234}$ at arbitrary $N$ with $X^I = 0$. The stabilizer of the trivial twisting background is $\Spin(7, \C) \ltimes \C^8 \times \C^\times$ and the ambient bulk geometry is of the form $\R^8 \times \C$. This twist is called $\Spin(7)$ twist.

\subsubsection{The holomorphic non-isotropic twist}
Now we move to twisting backgrounds in which the kinematical supercharge $\ekin \neq 0$. These require infinite $N$, since they must satisfy $\Theta \neq 0$. The minimal choice for such a twisting background arises when both $\ekin$ and $\edyn$ are pure spinors with their associated maximal isotropic subspaces perfectly aligned. To write representatives for such twisting backgrounds, we fix $\ekin = 1$ and $\edyn = 1$. 

The twisting supercharge $Q = \edyn \otimes w + \ekin \otimes w^\vee$ is of the holomorphic non-isotropic orbit described in~\cite{iib-nv}.\footnote{Here, the non-isotropic terminology refers to the IIB doublet direction. In this example, the representative uses $w + w^{\vee}$, which has nonzero self-pairing; the purely dynamical holomorphic twist, by contrast, uses $w$ which has vanishing self-pairing.} Since the two complex structures defined by $\edyn$ and $\ekin$ align, the geometric structure on the bulk is of type $\C^5$ and the stabilizer of this twisting supercharge is $\mathrm{SL}(5,\C) \ltimes \wedge^2 L$.

To make this into a twisting background of IKKT, we need to fix a non-commutative configuration for the bosonic matrices solving~\eqref{eq: tw-back-2}. From here, it immediately follows that $\Theta^{(2,0)} = 0$ and therefore, by Hermiticity, also $\Theta^{(0,2)} = 0$. For $\Theta^{(1,1)}$, we find the condition
\begin{equation}
    \Theta^i_i = \sum_{i = 1}^5 [Z^i , \bar Z_i] = 2.
\end{equation}
The complex dimension of the resulting brane worldvolume is classified by the rank of $\Theta^{(1,1)} \in \mathfrak{gl}(L)$ which can be $r = 1, \ldots, 5$. The corresponding brane extends along $r$ complex directions, furnishing a D$(2r-1)$-brane placed along a subspace $\CC^r \subset \CC^5$. The nonzero components of $\Theta^{(1,1)}$ determine the brane's constant worldvolume flux; recall that the flux of a brane is given by inverting $\Theta$ after restricting it to the worldvolume~\cite{Seiberg:1999vs}. This gives a constant two-form. Here, this form is of type $(1,1)$ so that we can view it as originating as the curvature of a holomorphic line bundle. Thus, the brane spectrum reproduces B-branes with worldvolumes of at least complex dimension one.

A minimal choice for a brane background in this twist is
\begin{equation}
     [Z^1 , \bar{Z_1}] = 2,
\end{equation}
corresponding to a D$1$ along $\C_{z_1} \subset \CC^5$. 

\subsubsection{The $\SU(3)$ twist} Keeping the dynamical twisting supercharge as a pure spinor, we can use the $\Spin(10,\C)$-action to fix $\edyn = 1$. If the kinematical twisting supercharge is also pure, the twisting background conditions imply that the dimension of  $L_{\varepsilon_{\mathrm{dyn}}} \cap L_{\varepsilon_\mathrm{kin}} \in  \{3,5\}$. If the dimension is $5$, the two pure spinors are perfectly aligned which recovers the holomorphic non-isotropic twist discussed above. For the dimension of the intersection being $3$, an explicit representative is given by $\ekin = -i \vartheta e^\vee_{12}$ for non-vanishing $\vartheta \in \R$. (For this representative, notice that contraction by $e_3, e_4, e_5$ annihilates both spinors, but contraction by $e_1, e_2$ fails to annihilate $\ekin$, hence $L_1 \cap L_{e^{\vee}_{12}} = \textrm{span}\left\lbrace e_3, e_4, e_5 \right\rbrace$, of dimension 3 as claimed).

The corresponding twist in type IIB supergravity is the $\mathrm{SU}(3)$ twist of~\cite{iib-nv}. The bulk spacetime takes the form $\R^4 \times \C^3$ and the stabilizer of such a twisting background is $(\mathrm{SL}(3, \C) \times \mathrm{SL}(2,\C)) \ltimes N_{15} \times \C^\times$, where $N_{15}$ is a nilpotent group of dimension 15.

The condition for a valid twisting background is
\begin{equation}
     \Theta \cdot 1 = i \vartheta e^\vee_{12} .
\end{equation}
This equation decomposes into zero-, two-, and four-form components. Here the four-form component is trivial, while the two-form piece fixes
\begin{equation}
    [Z^1, Z^2] = i \vartheta , \qquad [Z^i, Z^j] = 0 \quad \text{otherwise} ,
\end{equation}
and therefore, by Hermiticity, $[\bar Z_1, \bar Z_2] =  i \vartheta$. The zero-form component gives
\begin{equation}
    \tr(\Theta^{(1,1)}) = 0.
\end{equation}
Along $z_1, z_2$, we are thus in the situation of the example in~\S\ref{ssec: worldvolumes}. We split the $2 \times 2$ block of $\Theta^{(1,1)}$ in the directions $i,j = 1,2$ into its trace and its traceless part $\mathring{\Theta}^{(1,1)}$. An $\SU(2)$ rotation of $z_1, z_2$, which preserves the twisting supercharge, brings the latter to the form
\begin{equation}
    [Z^1, \bar Z_2] = [\bar Z_1, Z^2] = i \rho , \qquad \rho \in \R .
\end{equation}
The commutators of $X^1, \dots, X^4$ then generically have rank four, so that the worldvolume fills $\R^4_{x_1, \dots, x_4}$. If the trace vanishes, we find that the rank drops to two when $\rho = \pm \vartheta$. Then the brane is placed along the real plane $\R^2_{x_1, x_3}$ or $\R^2_{x_2, x_4}$. \footnote{More generally, writing $[Z^1, \bar Z_1] + [Z^2, \bar Z_2] = \tau$ for the trace of the block, the rank drops to two for $\rho^2 = \vartheta^2 + \tau^2/4$ and the plane is tilted away from the coordinate directions.}

In addition, we can have non-commutative directions of type $(1,1)$ in the remaining complex directions $z_3, z_4, z_5$, defining a brane along a complex subspace $\C^r \subset \C^3$ for $r = 0, 1, 2, 3$. Their trace has to be compensated by the trace along the directions $i,j=1,2$ so that the total trace over all five directions still vanishes. In total, we obtain a brane spectrum consisting of products:
\begin{itemize}
    \item[---] Along the four topological directions, either a brane along $\R^2 \subset \R^4$ or a space-filling brane $\R^4$.
    \item[---] Along the three holomorphic directions, branes supported on complex subspaces of dimension $r = 0, 1, 2, 3$.
\end{itemize}
This is the expected brane spectrum for a mixed A/B model on $\R^4 \times \C^3$: the brane along $\R^2$ is a Lagrangian A-brane, while the space-filling brane on $\R^4$ is coisotropic~\cite{KapustinOrlov}. To make this identification precise, note that the kinematical twisting supercharge $\ekin = -i \vartheta e^\vee_{12}$ equips the bulk with a holomorphic bivector. Inverting this bivector gives the two-form
\begin{equation}
    B + i \omega = \vartheta^{-1} \d z_1 \wedge \d z_2,
\end{equation}
whose real and imaginary parts define the B-field and symplectic form for the A-model directions. Explicitly, we have
\begin{equation}
    \omega = \vartheta^{-1} (\d x_1 \wedge \d x_4 + \d x_2 \wedge \d x_3) \qquad B = \vartheta^{-1} (\d x_1 \wedge \d x_3 - \d x_2 \wedge \d x_4) .
\end{equation}
Thus, for the two-dimensional branes arising at $\vartheta = \pm \rho$, we see that $\omega$ restricts to zero on the worldvolume and hence that these branes are indeed Lagrangian. Inverting the restriction of the bivector $\Theta$ to the worldvolume defines the flux $F$ of the brane. From~\eqref{eq: nc-real}, we can read off
\begin{equation}
    F = -\vartheta^{-1} \d x_1 \wedge \d x_3,
\end{equation}
so that the restriction of $F+B$ vanishes on the worldvolume as required.

For a space-filling brane with $\vartheta \neq \rho$, we read off the flux
\begin{equation}
    F = - \frac{2 \, \d x_1 \wedge \d x_3}{\vartheta + \rho} + \frac{2 \, \d x_2 \wedge \d x_4}{\vartheta - \rho}
\end{equation}
Thus, we immediately find that $(F + B) \wedge \omega = 0$. Further, expanding the terms explicitly, we also find $(F + B) \wedge (F + B) - \omega \wedge \omega = 0$. On the worldvolume, we can define adapted coordinates $w_a$ so that,
\begin{equation}
    F + B + i \omega = \frac{\d w_1 \wedge \d w_2}{\vartheta (\rho^2 - \vartheta^2)} , \qquad w_a = \rho z_a - \vartheta \bar z_a ,
\end{equation}
becomes a holomorphic-symplectic form on the worldvolume with respect to the complex structure defined by $(w_a , \bar w_a)$. In other words, this brane background indeed defines a space-filling coisotropic A-brane~\cite{KapustinOrlov}.

\subsubsection{The two $\mathrm{Sp}(4)$-twists}
We have two orbits where precisely one of the dynamical or the kinematical twisting supercharges is pure while the other one is impure. These two orbits correspond to the single orbit of $\mathrm{Sp}(4)$-twisting supercharges in the nilpotence variety of the type IIB supertranslation algebra~\cite{iib-nv}. These orbits are exchanged under the $\Z_2$ that swaps kinematical and dynamical supersymmetries and hence define distinct twisting backgrounds for the IKKT model.

The stabilizers of such twisting supercharges are $\Sp(4,\C) \ltimes N_{13} \times \C^\times$ and the background spacetime is of the form $\R^8 \times \C$.

First, let the dynamical supercharge be pure and fix $\varepsilon_\mathrm{dyn} = 1$. For the impure kinematical supercharge, we choose the representative $\ekin = -i \vartheta (e^\vee_{12} + e^\vee_{34}) =: -i \vartheta \kappa$, where we have defined $\kappa = (e^\vee_{12} + e^\vee_{34})$. We refer to this twist as the $\mathrm{Sp}(4)_A$-twist.\footnote{Our conventions are such that Sp$(4,\C)$ preserves an antisymmetric form on a four-dimensional complex vector space.}
The condition for a valid twisting background is
\begin{equation}
    \Theta \cdot 1 = i \vartheta \, \kappa .
\end{equation}
The four-form component of this equation is trivial, while the two-form piece fixes
\begin{equation}
    [Z^1, Z^2] = [Z^3, Z^4] = i \vartheta , \qquad [Z^i, Z^j] = 0 \quad \text{otherwise} ,
\end{equation}
and therefore also by Hermiticity $[\bar Z_1, \bar Z_2] = [\bar Z_3, \bar Z_4] = i \vartheta$. The zero-form component gives
\begin{equation}
    \tr(\Theta^{(1,1)}) = 0 .
\end{equation}
Each of the pairs $z_1, z_2$ and $z_3, z_4$ can thus be analyzed along the same lines as the previous example. Namely, we split the $2 \times 2$ blocks of $\Theta^{(1,1)}$ along the two pairs into their traces and traceless parts. An $\SU(2) \times \SU(2)$ rotation of $z_1, z_2$ and $z_3, z_4$, which preserves the twisting supercharge, brings the traceless parts to the form
\begin{equation}
    [Z^1, \bar Z_2] = [\bar Z_1, Z^2] = i \rho_1 , \qquad [Z^3, \bar Z_4] = [\bar Z_3, Z^4] = i \rho_2 , \qquad \rho_1, \rho_2 \in \R .
\end{equation}
Each pair then generically contributes four non-commutative real directions, and only two if $\rho_a^2 = \vartheta^2 + \tau_a^2/4$, where $\tau_a$ denotes the trace of the corresponding block. For simplicity, we set $\tau_1 = \tau_2 = 0$ below, so that this becomes $\rho_a = \pm \vartheta$. A brane along $\C_{z_5}$ requires $\tau_1 + \tau_2 \neq 0$, which tilts the worldvolumes as in the $\SU(3)$-twist.

The minimal choice $\rho_1 = \rho_2 = \vartheta$ gives a D$3$-brane along the real subspace $\R^4_{x_1, x_3, x_5, x_7}$, while keeping only one or neither of the pairs minimal gives a D$5$- or D$7$-brane with worldvolume dimensions six and eight. Terms off the diagonal $2 \times 2$ blocks lead to further solutions with the same worldvolume dimensions.

In addition, we can have a non-commutative direction of type $(1,1)$ along the remaining complex direction $z_5$, defining a brane along $\C_{z_5}$. Its trace has to be compensated by the traces along the two pairs so that the total trace over all five directions still vanishes. Thus, we obtain a brane spectrum consisting of products:
\begin{itemize}
    \item[---] Along the eight real directions, a brane along a real subspace $\R^4$ or $\R^6$, or a space-filling brane $\R^8$.
    \item[---] Along the complex direction $z_5$, either a point or a brane along $\C_{z_5}$.
\end{itemize}
This is the expected brane spectrum for a mixed A/B model on $\R^8 \times \C$: the brane along $\R^4$ is a Lagrangian A-brane, while the branes along $\R^6$ and $\R^8$ are coisotropic~\cite{KapustinOrlov}. This can be made precise using the same procedure as for the $\SU(3)$-twist. The kinematical twisting supercharge now equips the bulk with the holomorphic bivector $ \vartheta (\partial_{z_1} \wedge \partial_{z_2} + \partial_{z_3} \wedge \partial_{z_4})$ whose inverse gives the B-field and the symplectic form along the eight topological directions,
\begin{equation} \label{eq: sp4a-BiO}
    B + i \omega = \vartheta^{-1} (\d z_1 \wedge \d z_2 + \d z_3 \wedge \d z_4) ,
\end{equation}
In terms of the real variables, the non-commutative variables take the form:
\begin{equation}
    \begin{aligned}
        [X^1, X^3] & = \tfrac{i}{2} (\vartheta + \rho_1) , & [X^2, X^4] & = \tfrac{i}{2} (\rho_1 - \vartheta) , \\
        [X^5, X^7] & = \tfrac{i}{2} (\vartheta + \rho_2) , & [X^6, X^8] & = \tfrac{i}{2} (\rho_2 - \vartheta) .
    \end{aligned}
\end{equation}
Hence, $B$, $\omega$, and the flux $F$ decompose into the contributions of the two pairs, each of which is of the form discussed for the $\SU(3)$ twist. A pair with $\rho_a = \pm \vartheta$ contributes a plane on which $\omega$ and $F + B$ vanish, while a pair with $\rho_a \neq \pm \vartheta$ contributes a plane equipped with a holomorphic-symplectic form with respect to the complex structure defined by the coordinates $(w_a = \rho z_a - \vartheta \bar z_a)$. Thus, the D$3$-brane is a Lagrangian A-brane, and the D$7$-brane is a space-filling coisotropic brane. The D$5$-brane, say with $\rho_1 = \vartheta$ and $\rho_2 \neq \pm \vartheta$, is a coisotropic brane that is not space-filling. On the worldvolume $W = \R^2_{x_1, x_3} \times \R^4_{x_5, \dots, x_8}$,  both $\omega$ and $F+B$ vanish along the plane $\R^2_{x_1, x_3}$ and we find
\begin{equation}
    (F + B + i \omega)|_W = \frac{\d w_3 \wedge \d w_4}{\vartheta (\rho_2^2 - \vartheta^2)} , \qquad w_a = \rho_2 z_a - \vartheta \bar z_a ,
\end{equation}
which is a holomorphic-symplectic form along the directions $\R^4_{x_5, \dots, x_8}$ as required for coisotropic A-branes.

The other way round, we set $\varepsilon_\mathrm{kin} = 1$ and choose the impure spinor $\varepsilon_\mathrm{dyn} = 1 + \kappa$ for the dynamical twisting supercharge. We refer to this twist as the $\mathrm{Sp}(4)_B$-twist. In contrast to the previous cases, $\varepsilon_\mathrm{dyn}$ has a non-vanishing vector invariant $v^I = \varepsilon_\mathrm{dyn} \Gamma^I \varepsilon_\mathrm{dyn} \propto e_5$, so that~\eqref{eq: tw-back-2} sets $Z^5 = 0$, and by Hermiticity $\bar Z_5 = 0$. A brane along $z_5$ is therefore not allowed. The condition for a valid twisting background is
\begin{equation}
    1 + \Theta \cdot (1 + \kappa) = 0 ,
\end{equation}
whose zero-, two-, and four-form components are
\begin{equation}
    \begin{split}
        \text{0-form:} & \qquad 1 + \Theta^{(1,1)} \cdot 1 + \Theta^{(0,2)} \cdot \kappa = 0 , \\
        \text{2-form:} & \qquad \Theta^{(2,0)} \cdot 1 + \Theta^{(1,1)} \cdot \kappa = 0 , \\
        \text{4-form:} & \qquad \Theta^{(2,0)} \cdot \kappa = 0 .
    \end{split}
\end{equation}
The four-form component gives $[Z^1, Z^2] + [Z^3, Z^4] = 0$ (so also $[\bar Z_1, \bar Z_2] + [\bar Z_3, \bar Z_4] = 0$), so that $\Theta^{(0,2)} \cdot \kappa = 0$. The zero-form component then reduces to
\begin{equation}
    \tr(\Theta^{(1,1)}) = 2 .
\end{equation}
In contrast to the $\mathrm{Sp}(4)_A$-twist, the $(2,0)$ part is not fixed by the twisting supercharge. Instead, the two-form equation determines it in terms of $\Theta^{(1,1)}$; note that $\Theta^{(2,0)} = 0$ exactly when $\Theta^{(1,1)} \in \mathfrak{csp}(4) = \mathfrak{sp}(4) \oplus \C$, i.e.\ when $\Theta^{(1,1)}$ preserves $\kappa$ up to a multiple.

Let us first consider backgrounds with $\Theta^{(2,0)} = 0$. We can diagonalize $\Theta^{(1,1)}$ by a transformation in $\mathrm{USp}(4)$ and we therefore take
\begin{equation}
    [Z^i, \bar Z_i] = t_i , \qquad [Z^i, \bar Z_j] = 0 \quad \text{for } i \neq j .
\end{equation}
The trace condition says $t_1 + t_2 + t_3 + t_4 = 2$, while preserving $\kappa$ up to a multiple requires $t_1 + t_2 = t_3 + t_4$. Thus, we find
\begin{equation}
    t_1 + t_2 = t_3 + t_4 = 1 .
\end{equation}
A minimal solution is given by
\begin{equation}
    [Z^1, \bar Z_1] = [Z^3, \bar Z_3] = 1
\end{equation}
and describes a D3 brane. Turning on further $t_i$ gives D5- and D7-branes, for instance
\begin{equation}
    t = (1, 0, 1 - t_4, t_4) \qquad \text{and} \qquad t = (1 - t_2, t_2, 1 - t_4, t_4) ,
\end{equation}
with $t_2, t_4 \neq 0, 1$.

Turning on $\Theta^{(2,0)}$ tilts the worldvolumes of these branes but does not change the list of dimensions. For instance, a diagonal background with $t_1 + t_2 \neq t_3 + t_4$ has $[Z^1, Z^2] = - [Z^3, Z^4] = \tfrac12 \big( (t_3 + t_4) - (t_1 + t_2) \big)$. Thus, we obtain a brane spectrum consisting of products:
\begin{itemize}
    \item[---] Along the eight real directions, a brane along a real subspace $\R^4$ or $\R^6$, or a space-filling brane $\R^8$.
    \item[---] Along the complex direction $z_5$, only a point.
\end{itemize}
Along the eight real directions, the worldvolumes have the same dimensions as those of the Lagrangian and coisotropic A-branes of the $\mathrm{Sp}(4)_A$-twist, but all branes are now localised at points in $\C_{z_5}$. Since $\edyn$ is impure, however, the BV complexes of fluctuations around these backgrounds are acyclic (see also~\S\ref{sec: spin7}). In contrast to the $\mathrm{Sp}(4)_A$-twist, these branes therefore carry no perturbative worldvolume theory.

\subsubsection{The $\SU(4)$ twist}
Finally, we have twisting backgrounds where both the kinematical and the dynamical twisting supercharge are impure spinors. In that case, the square-zero condition requires
\begin{equation}
    \Gamma(\varepsilon_\mathrm{dyn} , \varepsilon_{\mathrm{kin}}) = 0.
\end{equation}
This is solved for example by $\varepsilon_\mathrm{dyn}= 1+ e^\vee_{1234}$ and $\varepsilon_\mathrm{kin} = 1 - e^\vee_{1234}$. Note that the two linear combinations $\ekin \pm \edyn$ are pure so that twists of this type correspond to the $\SU(4)$ twist of type IIB supergravity. Indeed, for our representative we have $Q= 1 \otimes (w + w^\vee) + e^\vee_{1234} \otimes (w - w^\vee)$ which matches the representative of~\cite{iib-nv}.

The stabilizer of such a twisting supercharge is $\mathrm{SL}(4,\C) \ltimes N_8$, where $N_8$ is a nilpotent group of dimension 8, and the background is of the form $\R^8 \times \C$. We note that, while the bulk theory depends topologically on the eight directions, the twisting supercharge does equip them with a Calabi--Yau structure, and hence also with a complex structure. This follows from the compact real form of the semisimple part of the stabilizer being the Calabi-Yau fourfold holonomy group $\SU(4)$.

Similarly to the case of the $\mathrm{Sp}(4)_B$-twist, this twist has a non-vanishing vector invariant along $e_5$ forcing $Z^5=0$ and $\bar Z_5 = 0$. The condition for the background then becomes
\begin{equation}
    1 - e^\vee_{1234} + \Theta \cdot (1 + e^\vee_{1234}) = 0.
\end{equation}
Expanding into zero-, two-, and four-form components as above, we find that this condition is equivalent to\footnote{We remark that these equations are an infinite $N$ version of the ADHM equations studied in~\cite{Nekrasov:2017cih}.}
\begin{equation} \label{eq: su4-branes}
    \sum_{i=1}^4 [Z^i, \bar Z_i] = 2 , \qquad [Z^i, Z^j] = \tfrac12 \epsilon^{ijkl} [\bar Z_k, \bar Z_l] .
\end{equation}
Let us first set $\Theta^{(2,0)} = 0$. We can then use $\SU(4)$ to diagonalize $\Theta^{(1,1)}$, so that solutions are parametrized by
\begin{equation}
    [Z^i, \bar Z_i] = t_i \qquad t_1 + \dots + t_4 = 2,
\end{equation}
with the brane extending along $z_i$ when $t_i \neq 0$. This gives D1-, D3-, D5-, and D7-branes sitting along subspaces of $\RR^8$ that preserve the complex structure induced by the twisting supercharge. The minimal choice
\begin{equation}
    [Z^1, \bar Z_1] = 2
\end{equation}
describes a D1-brane.

For $\Theta^{(2,0)} \neq 0$, note that~\eqref{eq: su4-branes} together with Hermiticity implies it is non-degenerate. Hence, we use an $\SU(4)$ rotation to bring it to the form of the $\mathrm{Sp}(4)_A$-twist,
\begin{equation}
    [Z^1, Z^2] = [Z^3, Z^4] = i \vartheta .
\end{equation}
Hence, such a brane generically extends along all eight topological directions. Similar to the $\mathrm{Sp}(4)$-twists discussed above, turning on $\Theta^{(1,1)}$ allows for special loci on which the worldvolume dimension of the brane drops to six or four. These are branes that do not preserve the complex structure on $\R^8$ induced by the twisting supercharge.

In summary, we thus find the following brane spectrum:
\begin{itemize}
    \item[---] Along the eight real directions, either a D1-, D3-, D5-, or D7-brane along a complex subspace $\C^r \subset \C^4$ for $\Theta^{(2,0)} = 0$, or a D3-, D5-, or D7-brane along a real subspace $\R^4$, $\R^6$, or $\R^8$ for $\Theta^{(2,0)} \neq 0$.
    \item[---] Along the holomorphic direction $z_5$, only a point.
\end{itemize}
As for the $\mathrm{Sp}(4)_B$-twist, the BV complexes of fluctuations around these backgrounds are acyclic so that the worldvolume theories that these branes carry are trivial in perturbation theory.

\subsubsection{Summary and comparison with type IIB supergravity} \label{sec: summary}
Table~\ref{tab: tw-backgrounds} summarizes the twisting supercharges of the IKKT matrix model together with the conditions for the non-commutative background.

\begin{table}[!htbp]
\centering
\small
\setlength{\tabcolsep}{5pt}
\renewcommand{\arraystretch}{1.5}
\begin{tabular}{l|l|l|l|l}
Twist & $\edyn$ & $\ekin$ & Geometry & Conditions on $\Theta$ \\
\hline
hol.\ iso. & $1$ & $0$ & $\C^5$ & $\Theta = 0$ \\
$\Spin(7)$ & $1 + e^\vee_{1234}$ & $0$ & $\R^8 \times \C$ & $\Theta = 0$ \\
\hline
hol.\ non-iso. & $1$ & $1$ & $\C^5$ & $\Theta^{(2,0)} = 0$, \ $\tr \Theta^{(1,1)} = 2$ \\
$\SU(3)$ & $1$ & $- i \vartheta e^\vee_{12}$ & $\R^4 \times \C^3$ & $\Theta^{(2,0)} = i \vartheta e^\vee_{12}$, \ $\tr \Theta^{(1,1)} = 0$ \\
$\mathrm{Sp}(4)_A$ & $1$ & $- i \vartheta \kappa$ & $\R^8 \times \C$ & $\Theta^{(2,0)} = i \vartheta \kappa$, \ $\tr \Theta^{(1,1)} = 0$ \\
$\mathrm{Sp}(4)_B$ & $1 + \kappa$ & $1$ & $\R^8 \times \C$ & \begin{tabular}[t]{@{}l@{}} $Z^5 = 0$, \ $\tr \Theta^{(1,1)} = 2$, \\ $\Theta^{(2,0)} = - \Theta^{(1,1)} \cdot \kappa$ \end{tabular} \\
$\SU(4)$ & $1 + e^\vee_{1234}$ & $1 - e^\vee_{1234}$ & $\R^8 \times \C$ & \begin{tabular}[t]{@{}l@{}} $Z^5 = 0$, \ $\tr \Theta^{(1,1)} = 2$, \\ $\Theta^{ij} = \tfrac12 \epsilon^{ijkl} \Theta_{kl}$ \end{tabular} \\
\end{tabular}
\caption{Twisting backgrounds for the IKKT model, where $\kappa = e^\vee_{12} + e^\vee_{34}$. Traces are taken over the spacetime indices. The first two lines exist at arbitrary $N$; all other backgrounds require infinite $N$.}
\label{tab: tw-backgrounds}
\end{table}

In the twists with $\ekin \neq 0$, solving for $\Theta$ leads to the supports and fluxes of branes summarized in Table~\ref{tab: brane-spectra}. We will determine the theories of fluctuations on their worldvolumes in \S \ref{sec: twists}.
\begin{table}[!htbp]
\centering
\small
\setlength{\tabcolsep}{5pt}
\renewcommand{\arraystretch}{1.5}
\begin{tabular}{l|l|l|l|l}
Twist & Geometry & Topological part & Holomorphic part & Minimal brane \\
\hline
hol.\ non-iso. & $\C^5$ & --- & $\C^r$, \ $r = 1, \dots, 5$ & D1 on $\C$ \\
$\SU(3)$ & $\R^4 \times \C^3$ & $\R^2$, \ $\R^4$ & $\C^r$, \ $r = 0, \dots, 3$ & D1 on $\R^2$ \\
$\mathrm{Sp}(4)_A$ & $\R^8 \times \C$ & $\R^4$, \ $\R^6$, \ $\R^8$ & point, \ $\C$ & D3 on $\R^4$ \\
$\mathrm{Sp}(4)_B$ & $\R^8 \times \C$ & $\R^4$, \ $\R^6$, \ $\R^8$ & point & D3 on $\R^4$ \\
$\SU(4)$ & $\R^8 \times \C$ & $\C^r$, \ $r = 1, \dots, 4$ \quad ($\Theta^{(2,0)} = 0$) & point & D1 on $\C$ \\
 & & $\R^4$, \ $\R^6$, \ $\R^8$ \quad ($\Theta^{(2,0)} \neq 0$) & & \\
\end{tabular}
\caption{Brane spectra of the twisting backgrounds with $\ekin \neq 0$. 
For the block-diagonal representatives, each brane is the product of a worldvolume along the topological directions and one along the holomorphic directions; general backgrounds give tilted branes of the same dimensions.
$\C^r$ denotes subspaces that are complex with respect to the complex structure induced by $L$.}
\label{tab: brane-spectra}
\end{table}

Figure~\ref{fig:hasse-ikkt} shows the orbit stratification of the admissible twisting supercharges in the IKKT model. Each orbit is labeled by the (real form) of the semisimple part of the stabilizer of the corresponding twisting supercharge, the ten-dimensional ambient geometry, and the simplest matrix background that realizes the twist.

Figure~\ref{fig:hasse-ikkt-iib} illustrates the comparison of these orbits with orbits of the nilpotence variety $Y_{IIB}$ that forms the moduli space of twists for type IIB supergravity in a flat background.
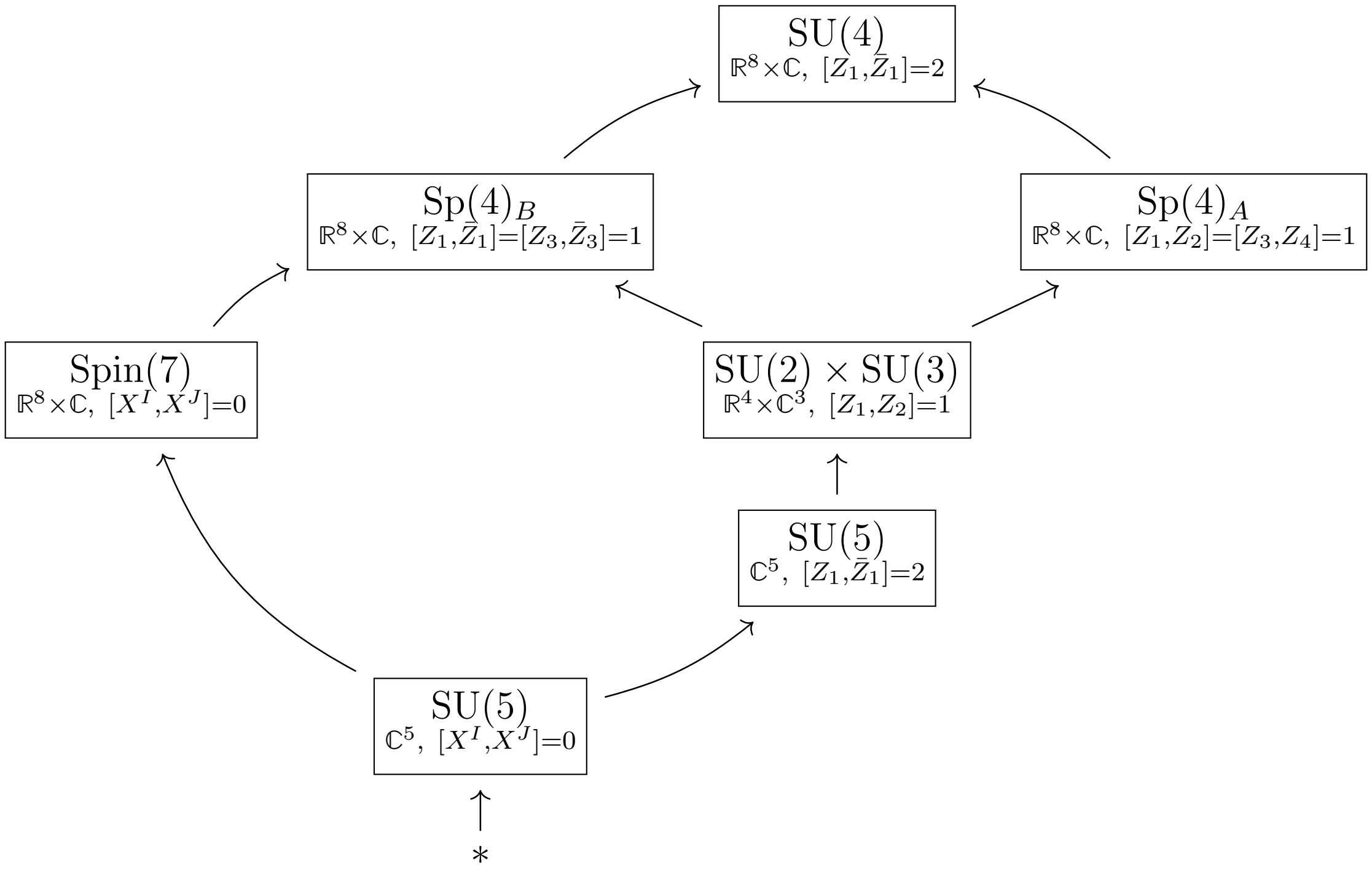
\begin{figure}[!htbp]
	\centering
	\begin{tikzcd}[row sep=1.1em, column sep=0.4em, every node/.style={inner sep=3pt}]
	& & {\boxed{\begin{gathered}\SU(4)\\[-0.7em] {\scriptstyle \RR^8\times\CC,\ [Z_1,\bar Z_1]=2}\end{gathered}}} & & \\
	& {\boxed{\begin{gathered}\mathrm{Sp}(4)_B\\[-0.7em] {\scriptstyle \RR^8\times\CC,\ [Z_1,\bar Z_1]=[Z_3,\bar Z_3]=1}\end{gathered}}}
	\arrow[ur, bend left=12] & & {\boxed{\begin{gathered}\mathrm{Sp}(4)_A\\[-0.7em] {\scriptstyle \RR^8\times\CC,\ [Z_1,Z_2]=[Z_3,Z_4]=1}\end{gathered}}}
	\arrow[ul, bend right=12] & \\
	{\boxed{\begin{gathered}\Spin(7)\\[-0.7em] {\scriptstyle \RR^8\times\CC,\ [X^I,X^J]=0}\end{gathered}}}
	\arrow[ur, bend left=12] & & {\boxed{\begin{gathered}\SU(2) \times \SU(3)\\[-0.7em] {\scriptstyle \RR^4\times\CC^3,\ [Z_1,Z_2]=1}\end{gathered}}}
	\arrow[ul] \arrow[ur] & & \\
	& & {\boxed{\begin{gathered}\SU(5)\\[-0.7em] {\scriptstyle \CC^5,\ [Z_1,\bar Z_1]=2}\end{gathered}}}
	\arrow[u] & & \\
	& {\boxed{\begin{gathered}\SU(5)\\[-0.7em] {\scriptstyle \CC^5,\ [X^I,X^J]=0}\end{gathered}}}
	\arrow[ur, bend right=12]
	\arrow[uul, bend left=20] & & & \\
	& * \arrow[u] & & &
	\end{tikzcd}
	\caption{Orbit stratification of admissible IKKT twisting supercharges. Each node includes the semisimple part of the stabilizer of the twisting supercharge, ambient geometry, and a minimal matrix background. An arrow $A \rightarrow B$ means that $A$ lies in the closure of $B$, so that representatives of $A$ arise as limits of representatives of $B$. The symbol $*$ denotes the zero supercharge included as a limiting point. The two purely dynamical twists admit realizations at any finite $N$, while the remaining types require $N = \infty$. }
	\label{fig:hasse-ikkt}
\end{figure}
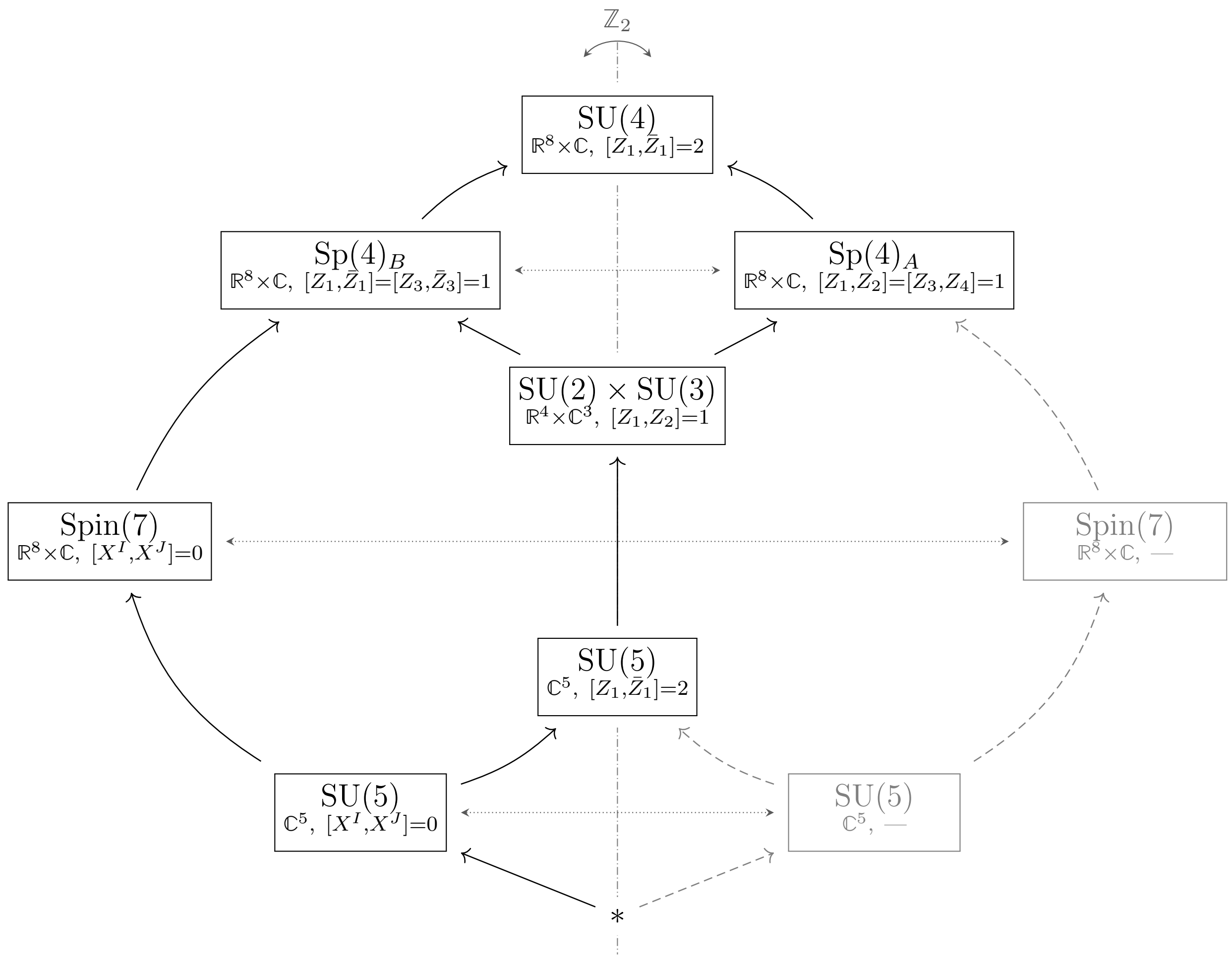
\begin{figure}[!htbp]
	\centering
    \newlength{\samesizewd}
    \newcommand{\samesize}[2]{\settowidth{\samesizewd}{$\scriptstyle #1$}\makebox[\samesizewd]{$\scriptstyle\vphantom{#1}#2$}}
	\begin{tikzcd}[row sep=1.1em, column sep=-0.6em, every node/.style={inner sep=3pt},
		cells={nodes={fill=white}},
		execute at end picture={
			\begin{scope}[on background layer]
				\draw[gray, densely dash dot]
					($(\tikzcdmatrixname-1-3.north)+(0,0.55)$) --
					($(\tikzcdmatrixname-7-3.south)+(0,-0.25)$);
			\end{scope}
			\draw[<->, >=stealth, gray!70!black]
				($(\tikzcdmatrixname-1-3.north)+(-0.45,0.35)$) to[bend left=50]
				node[above, font=\footnotesize] {$\mathbb{Z}_2$}
				($(\tikzcdmatrixname-1-3.north)+(0.45,0.35)$);
			\begin{scope}[on background layer]
				\draw[<->, >=stealth, densely dotted, gray!70!black]
					(\tikzcdmatrixname-2-2) -- (\tikzcdmatrixname-2-4);
				\draw[<->, >=stealth, densely dotted, gray!70!black]
					(\tikzcdmatrixname-4-1) -- (\tikzcdmatrixname-4-5);
				\draw[<->, >=stealth, densely dotted, gray!70!black]
					(\tikzcdmatrixname-6-2) -- (\tikzcdmatrixname-6-4);
			\end{scope}
		}]
	& & {\boxed{\begin{gathered}\SU(4)\\[-0.7em] {\scriptstyle \RR^8\times\CC,\ [Z_1,\bar Z_1]=2}\end{gathered}}} & & \\
	& {\boxed{\begin{gathered}\mathrm{Sp}(4)_B\\[-0.7em] {\scriptstyle \RR^8\times\CC,\ [Z_1,\bar Z_1]=[Z_3,\bar Z_3]=1}\end{gathered}}}
	\arrow[ur, bend left=12] & & {\boxed{\begin{gathered}\mathrm{Sp}(4)_A\\[-0.7em] {\scriptstyle \RR^8\times\CC,\ [Z_1,Z_2]=[Z_3,Z_4]=1}\end{gathered}}}
	\arrow[ul, bend right=12] & \\
	& & {\boxed{\begin{gathered}\SU(2) \times \SU(3)\\[-0.7em] {\scriptstyle \RR^4\times\CC^3,\ [Z_1,Z_2]=1}\end{gathered}}}
	\arrow[ul] \arrow[ur] & & \\
	{\boxed{\begin{gathered}\Spin(7)\\[-0.7em] {\scriptstyle \RR^8\times\CC,\ [X^I,X^J]=0}\end{gathered}}}
	\arrow[uur, bend left=15] & & & & {\color{gray}\boxed{\begin{gathered}\Spin(7)\\[-0.7em] \samesize{\RR^8\times\CC,\ [X^I,X^J]=0}{\RR^8\times\CC,\ \text{---}}\end{gathered}}}
	\arrow[uul, bend right=15, gray, dashed] \\
	& & {\boxed{\begin{gathered}\SU(5)\\[-0.7em] {\scriptstyle \CC^5,\ [Z_1,\bar Z_1]=2}\end{gathered}}}
	\arrow[uu] & & \\
	& {\boxed{\begin{gathered}\SU(5)\\[-0.7em] {\scriptstyle \CC^5,\ [X^I,X^J]=0}\end{gathered}}}
	\arrow[ur, bend right=12]
	\arrow[uul, bend left=20] & & {\color{gray}\boxed{\begin{gathered}\SU(5)\\[-0.7em] \samesize{\CC^5,\ [X^I,X^J]=0}{\CC^5,\ \text{---}}\end{gathered}}}
	\arrow[ul, bend left=12, gray, dashed]
	\arrow[uur, bend right=20, gray, dashed] & \\
	& & * \arrow[ul] \arrow[ur, gray, dashed] & &
	\end{tikzcd}
	\caption{The nine nonzero IIB supercharge orbits under $\Spin(10, \C) \times \SO(2, \C)$. The seven black nodes admit IKKT realizations, while the two grey nodes are purely kinematical, indicating twists which are disallowed in the matrix model. Solid and dashed arrows denote closures as described in Figure \ref{fig:hasse-ikkt}. Horizontal double arrows denote pairs exchanged by the disconnected $\ZZ_2 \subset \mathrm{O}(2, \C)$.}
	\label{fig:hasse-ikkt-iib}
\end{figure}
\newpage
The moduli space of twists for type IIB supergravity in a flat background was described in~\cite{iib-nv}. It is $Y_{IIB}$, the space of odd square-zero elements in the type IIB super Poincar\'e algebra together with the orbit stratification coming from $\mathrm{Spin}(10,\C) \times \mathrm{O}(2,\CC)$, where the latter factor arises as an R-symmetry. \footnote{We remark that there are some differences expected between twisted type IIB supergravity and twisted type IIB string theory. For example, the holomorphic twist of the IIB supergravity is described by minimal BCOV theory, whereas the open--closed construction returns full BCOV theory. However, the latter differs from the former only by additional nonpropagating fields, which are expected to encode the twist of massive string states but do not influence our discussion; see \cite[Sec.~2.4]{tw-bfss-ikkt} and \cite[Secs.~4--5]{CostelloLi}.} To compare this situation with the twists of the IKKT model, we refine the orbit structure of $Y_{IIB}$ by restricting the R-symmetry group to $\mathrm{SO}(2,\C) \subset \mathrm{O}(2,\C)$. Then, twists of the IKKT model at infinite $N$ realize twisting supercharges from all orbits where $\edyn \neq 0$. However, we find that the orbit of $\mathrm{Sp}(4)$-equivariant twists of $Y_{IIB}$ splits into two orbits in the IKKT model. Further, the orbit of holomorphic isotropic twists and that of $\Spin(7)$-twists have two components in $Y_{IIB}$ that are exchanged under $\Z_2$. The IKKT model only realizes representatives from one of these components. Figure \ref{fig:hasse-ikkt-iib} illustrates the relation between twists of IKKT and $Y_{IIB}$.

\section{A survey of twisted IKKT models} \label{sec: twists}
In this section, we describe the different twisted theories arising as fluctuations around the twisted backgrounds classified above. We do this by building on the holomorphic isotropic twist computed in~\cite{tw-bfss-ikkt}. This is the minimal twist of the IKKT model, and all other twists can be obtained as further deformations from it. The surviving fluctuation fields and their linearized kinetic operator give the basic data of a BV complex; concretely, expanding the cubic action of the minimal twist around a background field configuration $\widehat \alpha$ determines the linearized differential \(D=[\widehat\alpha,-]\), together with the interaction inherited from the matrix product and the cyclic trace. 

We compute the resulting BV complexes and identify them with Dolbeault or de Rham complexes on the brane worldvolume. In particular, a noncommutative matrix background transmutes the adjoint actions of background matrices into, at infinite $N$, derivatives along the emergent brane in spacetime. Finally in~\S\ref{sec: spin7}, for the \textrm{Spin}$(7)$ twist whose BV complex is perturbatively trivial (or \textit{acyclic}) we discuss the relation between evaluating its BV path integral and supersymmetric localization.

\subsection{The minimal twist}
In~\cite{tw-bfss-ikkt}, the minimal twist of the IKKT model was computed and its BV complex identified as
\begin{equation} \label{eq: bv-min-tw}
    A_N = \CC[\theta^1, \dots, \theta^5] \otimes \mathfrak{gl}_N
\end{equation}
with the BV action being the cubic expression $ \frac{1}{3}\int \d^5 \theta \tr(\alpha [\alpha , \alpha])$. 

Here, the $\theta^i$ are anticommuting variables, with $\mathbb{C}[\theta^1, \ldots, \theta^5]$ therefore denoting the exterior algebra. The superfield $\alpha$ packages the BV fields of Table \ref{tbl:minimaltwist} as its coefficients. The expansion begins $\alpha = c + \theta_i Z^i + \ldots$. The Berezin integral $\int d^5\theta$ extracts the coefficient of $\theta^1 \ldots \theta^5$, and the bracket notation combines matrix multiplication with the Grassmann multiplication of the $\theta^i$. In the trivial background the action is exactly cubic, so the linearized BV differential vanishes. We will see that expanding about the backgrounds below generates a nonzero differential for the fluctuations.

Explicitly, the field $\alpha \in A_N$ decomposes into fields of the IKKT model as follows.
\begin{table}[h]
\centering
\renewcommand{\arraystretch}{1.5}
\begin{tabular}{|c|c|c|c|c|c|}
\hline
$(\theta^i)^0$ & $(\theta^i)^1$ & $(\theta^i)^2$ & $(\theta^i)^3$ & $(\theta^i)^4$ & $(\theta^i)^5$ \\
\hline
$c$ & $Z^i$ & $\psi^{(2)}$ & $\psi^{*(3)}$ & $Z^*_i$ & $c^*$ \\
\hline
\end{tabular}
\caption{Fields of minimally twisted IKKT.}\label{tbl:minimaltwist}
\end{table}
For what follows, let us briefly review some aspects of the minimally twisted theory and its closed string dual.

\subsubsection{Residual symmetries}
Twisting deforms the type IIB super Poincar\'e algebra to the differential graded, or dg, Lie algebra $(\fp, [Q,-])$. In particular, to find the symmetries that survive the twist, we have equipped the super Poincar\'e algebra with the differential $d_Q = [Q, -]$. A surviving symmetry commutes with $Q$, while a transformation of the form $[Q, -]$ acts trivially on $Q$-cohomology. The residual symmetries are therefore the cohomology of this differential, which we denote by $\mathfrak{p}_Q$\footnote{In general, this residual symmetry algebra may be an $L_{\infty}$ algebra, rather than a Lie algebra. The higher brackets can be computed via homotopy transfer; see \cite{Jonsson:2024uyr, HPR}.}. 

The dg Lie algebra has the following structure:
\begin{equation}
    \begin{tikzcd}[row sep = tiny]
    \wedge^2 L & \wedge^4 L^\vee \otimes \langle w^{\vee} \rangle\arrow[r] & L \\
    \mathfrak{sl}(5) & \wedge^4 L^\vee \otimes \langle w \rangle & L^\vee \\
    & \wedge^2 L^\vee \otimes \langle w^{\vee} \rangle &  \\
    \wedge^2 L^\vee \arrow[r] & \wedge^2 L^\vee \otimes \langle w \rangle \\
    \CC \arrow[rd] & \wedge^0 L^\vee \otimes \langle w^{\vee} \rangle \\
    \CC \arrow[r] & \wedge^0 L^\vee \otimes \langle w \rangle
	\end{tikzcd}
\end{equation}
Here, the first column describes the decomposition of $\mathfrak{so}(10,\C) \oplus \mathfrak{so}(2,\C)$; note that the two abelian subalgebras denoted by $\C$ arise as the center in $\mathfrak{gl}(5) \subset \mathfrak{so}(10,\C)$ and from the R-symmetry $\mathfrak{o}(2,\C) \cong \CC$. The arrows show which generators are paired by $d_Q$. We can read off the cohomology by removing such paired generators to find the residual supersymmetries that act on the twisted theory:
\begin{equation} \label{eq: res-iib}
    (\mathfrak{sl}(5) \oplus \wedge^2 L \oplus \CC) \oplus \Pi ( (\wedge^0 L^\vee \oplus \wedge^2 L^\vee) \otimes \langle w^\vee \rangle \oplus \wedge^4 L^\vee \otimes \langle w \rangle) \oplus L^\vee.
\end{equation}
Note that the odd part features the four form component of the dynamical supersymmetries $\wedge^4 L^\vee \otimes \langle w \rangle$ together with the zero- and two-form components of the kinematical supersymmetries. The bracket between two odd elements takes the two-form piece of the kinematical supercharge and the four-form dynamical supercharge to give a holomorphic translation
\begin{equation}
    \wedge^2 L^\vee \otimes \langle w \rangle^\vee \times \wedge^4 L^\vee \otimes \langle w \rangle \longrightarrow L^\vee \qquad (\ekin^{(2)} , \edyn^{(4)}) \mapsto \ekin^{(2)} \vee \Omega^{-1} \edyn^{(4)}.
\end{equation}
Here, we have fixed the volume form $\Omega$ on $L$; in components we take $\Omega$ to be the totally antisymmetric symbol $\Omega^{ijklm} = \epsilon^{ijklm}$. This identifies a four-form $\eta$ with a vector $u$ that satisfies $\iota_u \Omega = \eta$; we denote this identification by the symbol $\Omega^{-1}$.  The $\vee$ symbol above denotes contraction. 

Fixing the twisting supercharge in one input of the ternary extension leaves a bracket on the two remaining inputs. In this way, the three-ary extension of \S \ref{sec: classif} produces a central binary bracket between a residual dynamical supercharge and a holomorphic translation. The result is 
\begin{equation} \label{eq: central-ext}
    \wedge^4 L^\vee \otimes \langle w \rangle \otimes L^\vee \mapsto \CC \qquad (\edyn^{(4)}, v) \mapsto \Omega^{-1}(\edyn) \vee v.
\end{equation}
The residual dynamical supersymmetry $\edyn^{(4)}$ acts on the fields according to the following formulas.
\begin{equation}
\begin{array}{rcl}
\delta c &=& \epsilon_{ijklm} \edyn^{(4) ijkl} Z^m \\[1ex]
\delta Z^i &=& \edyn^{(4)ijkl} \psi^{(2)mn} \varepsilon_{jklmn} \\[1ex]
\delta \psi^{(2)ij} &=& \epsilon_{klmno} \edyn^{(4)lmno} \psi^{*(3)ijk} \\[1ex]
\delta \psi^{*(3)ijk} &=& \edyn^{(4)ijkl} Z^*_l \\[1ex]
\delta Z^*_i &=& \epsilon_{ijklm} \edyn^{(4)jklm} c^* \\[1ex]
\delta c^* &=& 0
\end{array}
\end{equation}
In terms of the superfield $\alpha$, this transformation rule is summarized as
\begin{equation}
    \delta \alpha = \epsilon_{ijklm} \partial_{\theta^m} \edyn^{(4) ijkl} \alpha = \edyn^{(4)} \partial_\theta \alpha . 
\end{equation}
The residual kinematical supersymmetry acts on the trace part of $\psi^{(2)}$ by the usual shift
\begin{equation}
    \delta \psi^{(2)} = \ekin^{(2)} \id,
\end{equation}
as well as holomorphic translations that shift $Z^i$.

The full twisted BV action including the kinematical supersymmetries, translations and the central extension reads
\begin{equation}
    S_{BV} = \int \d^5 \theta \left(  \tr(\alpha [\alpha, \alpha] + \star \alpha \: \edyn^{(4)} \partial_\theta \alpha) \right) + \ekin^{(2)} \tr(\psi^{*(3)}) + a^i \tr(Z^*_i) + b^* a_i \edyn^{(4)jklm} \epsilon_{ijklm} + b \tr(c^*)
\end{equation}
where we denote by $\star \alpha$ the dual of $\alpha$ under the wedge pairing, i.e. its expansion in $\theta$-degree is $\star \alpha = c^* + \theta^i Z^*_i + \dots$.

\subsubsection{Large-$N$ duality and backreaction}
There is a standard relation in open-closed string field theory that closed-string fields are identified with cyclic deformations of the open-string action (from the worldsheet perspective, these are disc amplitudes with a single closed string insertion in the interior and arbitrarily many open string insertions on the boundary). Such deformations can be computed by cyclic cohomology \cite{Moeller:2010mh} (see also \cite{Movshev:2009ba}). With this method, one can reconstruct the closed string state space directly from the open string sector. This can be made particularly tractable in the twisted setting, where the large-$N$ holomorphic twist of the matrix model furnishes the open string field theory\footnote{In the twisted setting, the D-instanton open string field theory is equivalent to the IKKT action up to a nonlinear field redefinition. In the untwisted setting however, the D-instanton OSFT retains $\alpha'$ corrections. The field theory limit of the D-instanton open string effective action and the relation to the matrix model proposal were studied in \cite{Ooguri:1998pf}.}, while the closed string field theory is described by BCOV theory \cite{BCOV, CLBCOV1} on $\mathbb{C}^5$. We computed this cyclic cohomology at infinite $N$ and found that the large-$N$ single-trace complex reproduces the BCOV field complex and its classical interaction brackets \cite{tw-bfss-ikkt}.

For comparison with IIB string theory, we use the BCOV description and residual symmetry map of \cite{CostelloLi}. The BV space of fields in BCOV theory is given as (formal power series in) polyvector fields
\begin{equation}
    \mathrm{PV}^{p, q}(\mathbb{C}^5)= \Omega^{(0, q)}(\mathbb{C}^5, \wedge^p T^{1, 0}),
\end{equation} which are $(0, q)$-forms valued in holomorphic $p$-vector fields. We use $\mathrm{PV}^{p, q}(\mathbb{C}^5)[[t]]$ to denote formal power series in an auxiliary parameter $t$ of cohomological degree 2.

The action of the residual symmetries on BCOV theory is given by an $L_\infty$ map into the BCOV field space
\begin{equation}
    \hat{\fp}_Q \rightsquigarrow \mathrm{PV}^{\bu,\bu}(\CC^5)[\![t]\!],
\end{equation}
where $\hat{\fp}_Q$ denotes the residual supersymmetry algebra including the central extension described above.
Explicit entries of the map on basis elements include:
\begin{equation} \label{eq: res-bcov}
    \edyn^{(4)} \mapsto z_i \in \mathrm{PV}^{0,0} \qquad \ekin^{(2)} \mapsto \partial_{z_i} \wedge \partial_{z_j} \in \mathrm{PV}^{2,0} \qquad \ekin^{(0)} \mapsto 0.
\end{equation}
An expectation value for the four-form component of the dynamical supersymmetry transformation therefore gives a linear bulk superpotential $W = \lambda^i z_i$, whose effect is to add the bracket with $W$ to the BCOV theory's BV differential. This renders the theory \textit{acyclic}, which means that the resulting linearized BV complex has zero cohomology. On the other hand, adding a kinematical two-form supercharge corresponds to adding a bivector deformation. The zero-form part of the kinematical supersymmetry does not deform the bulk theory at all, since it is in the kernel of this map.

We remark that (the homotopy transfer of) the central extension as given in~\eqref{eq: central-ext} corresponds to the elementary bracket $[\partial_{z^i}, z^j] = \delta^j_i$ in BCOV theory.

It is worth pointing out that the background in which the holographically dual BCOV theory lives is generally non-trivial due to backreaction. Indeed, the closed string background depends on the brane's disc one-point function, which has one closed-string insertion but no open-string insertions, cf. \cite[Sec.~8]{CostelloLi}. For our stack of $N$ D$(-1)$s in the minimal twist, the field in BCOV theory that is sourced sits in $\mathrm{PV}^{0,4}$ which, using the holomorphic volume form, corresponds to the $(5,4)$ component of the D-instantons' magnetic nine-form flux. Its class in $H^{0,4}(\C^5 \setminus 0)$ is detected by the flux through a nine-sphere linking the stack. The dual BCOV theory is then to be expanded around this background which shifts the linearized BV differential by the Schouten bracket with this flux; for $N$ coincident point-branes, its Chan--Paton factor is $\operatorname{tr}(1_N)=N$, resulting in the usual backreaction of a brane stack in twisted holography~\cite{Costello:2018zrm, Costello:2020jbh}\footnote{Here, keeping track of the magnetic flux sourced by a D-instanton stack does not distinguish the full background from its near-horizon limit. In Euclidean conventions $e^\phi=H=h+q_{\mathrm D}/r^8$ and $\chi=-H^{-1}$, with
Einstein-frame Hodge star $*_E$, one has
$F_9=e^{2\phi}*_E d\chi=*_E dH$.
This is independent of $h$ at fixed charge \cite{Ciceri:2025maa}. This fact is paralleled in twisted holography, where the twisted backreacted geometry is unchanged by a near-horizon limit \cite{Costello:2018zrm, Costello:2020jbh}.}.

Below we study twists of the IKKT model that feature brane backgrounds. In that situation, the D-instantons have dissolved into higher dimensional D$p$ branes, and their D$(-1)$-charge is smeared out to a density across the worldvolume of the brane~\cite{Aoki:1999vr, Chepelev:1997vx}. If the D$(-1)$s dissolve into a stack of $k$ branes, the stack thus carries $k$ units of D$p$ brane charge, the D$(-1)$-charge density, and possibly other charges corresponding to other lower-dimensional dissolved branes. All these charges act as sources for the closed string fields. We leave a careful treatment of this backreaction to future work. Here, we remark that for D$(-1)$-branes dissolving into B-branes with affine worldvolume $\CC^r \subset \C^5$, the smeared D-instanton charge is cohomologically trivial in the complement $H^{0,4}(\CC^5 \setminus \CC^r) = 0$ for $ 1 \leq r \leq 4$ (although the boundary can retain dependence on the D-instanton charge). The D$p$-brane charge, in contrast, sources a non-trivial class even in the complement $H^{0,4-r}(\C^5 \setminus \C^r)$~\cite[Sec.~8]{CostelloLi}.

We also remark that the twisted correspondence includes protected components of the familiar IKKT vertex operators \cite{Kitazawa:2002vh,Kitazawa:2008mp}. This is reassuring, since in the untwisted setting these vertex operators couple the matrices to the IIB supergravity multiplet, hence protected components should remain in the cyclic complex which governs the closed string deformations of the twisted matrix model. For example, consider the complex axion--dilaton vertex
\[
  \mathcal V_\Phi(k)=\operatorname{tr}e^{\mathrm{i}k_I X^I}.
\]
Choose a complex momentum satisfying
$k_I\Gamma^I\varepsilon_{\rm dyn}=0$, so that it lies in the
isotropic momentum plane selected by the holomorphic twist and
$k_I X^I=k_iZ^i$. Our transformation $Q_{\rm BV}Z^i=[c,Z^i]$ implies
\[
  Q_{\rm BV}\mathcal V_\Phi(k)
  =\operatorname{tr}[c,e^{\mathrm{i}k_iZ^i}]=0.
\]
Here, we understand the exponential as a formal generating series of polynomial observables. The nonconstant Taylor coefficients are nonexact: to see this, restrict all BV fields to
$\mathbb C E_{11}\subset\mathfrak{gl}_N$, where $E_{11}$ is a rank-one diagonal projector. This makes the entire differential vanish, but the scalar traces remain nonzero. There are similarly fermionic classes
\[
  V_n^{ij}
  =\operatorname{tr}\bigl(\psi^{(2)ij}(Z^i)^n\bigr),
  \qquad i\ne j,\quad n\geq0.
\]
Their BV variation is proportional to
$\operatorname{tr}([Z^i,Z^j](Z^i)^n)$ which vanishes by cyclicity, while the same abelian restriction used above establishes their nonexactness. These operators occur as Taylor coefficients of dilatino vertex components. Holomorphic components of the two-form vertex operator formed from the $B$-field and the RR 2-form in the bulk also survive the twist. In the twisted IKKT model, the tensor vertex operators include terms proportional to the antifield $\psi^{*(3)}$, which cancels terms proportional to the equations of motion in the supersymmetry variation of the physical vertex.
\footnote{The vertex operators above are components of the BPS multiplets $\mathcal{B}_n$ into which~\cite{Ciceri:2025maa, Ciceri:2025wpb} organize the single-trace operators of the IKKT model. We expect the minimal twist to contain the holomorphic components of these BPS multiplets with respect to the complex structure selected by the twisting supercharge.}

\subsection{Further twists of minimally twisted IKKT}\label{sec:furthertwists}
Every twist can be obtained as a further deformation of the minimal twist described in the previous subsection. Let us therefore discuss how the non-commutative backgrounds descend to the minimal twist and how they deform the D-instanton worldvolume theory\footnote{We note that in the untwisted model, long-distance supercurrent interactions at one loop around noncommutative brane backgrounds have been matched to gravitino and dilatino exchange in ten-dimensional supergravity \cite{Kimura:2000ur}; it would be interesting to spell out the twisted version of this comparison.}.

At infinite $N$, we identify $\mathfrak{gl}_\infty$ as a subalgebra of the endomorphisms $\End(\cH)$ of an infinite dimensional vector space $\cH$. In a background featuring non-commutative directions 
\begin{equation}
    [X^I, X^J] = i \vartheta^{IJ} \id \qquad \text{with} \qquad \rank(\vartheta) = 2r,
\end{equation}
$\cH$ is a module for the Weyl algebra in $r$ Heisenberg pairs. We identify its endomorphisms as
\begin{equation}
    \C[x_1, \dots, x_{2r}] \otimes \mathfrak{gl}_k,
\end{equation}
where $x_1, \dots, x_{2r}$ are linear coordinates on the worldvolume $\im(\vartheta)$. They are equipped with the Moyal product
\begin{equation} \label{eq: moyal}
    \begin{aligned}
        (f \star_\vartheta g)(x) & = \exp \Big( \frac{i}{2} \, \vartheta^{IJ} \frac{\partial}{\partial x_I} \frac{\partial}{\partial x'_J} \Big) f(x) \, g(x') \Big|_{x' = x} \\
        & = f g + \frac{i}{2} \, \vartheta^{IJ} \, (\partial_I f) \, (\partial_J g) - \frac{1}{8} \, \vartheta^{IJ} \vartheta^{KL} \, (\partial_I \partial_K f ) \, (\partial_J \partial_L g) + \dots
    \end{aligned}
\end{equation}
Correspondingly, the space of fields for the holomorphic isotropic twist at infinite $N$ in such a background takes the form
\begin{equation} \label{eq: nc-fields}
     \CC[\theta^1, \dots, \theta^5] \otimes \CC[x_1, \dots, x_{2r}] \otimes \mathfrak{gl}_k .
\end{equation}
The adjoint action of the background translates to a derivative $[X^I, -] = i \vartheta^{IJ} \partial_{x^J}$. 

Recall that brane backgrounds were classified by $\Theta \in \wedge^2 V$ with components leading to type $(2,0)$, $(1,1)$, and $(0,2)$ corresponding to the commutators $[Z^i, Z^j]$, $[Z^i, \bar Z_j]$ and $[\bar Z_i, \bar Z_j]$ respectively. Note that only $Z^i$ descends to a field of the minimal twist, while the action of $\bar Z_j$ is trivialized. Hence, a non-commutative background descends to the twisted theory by equipping~\eqref{eq: nc-fields} with the differential
\begin{equation} \label{eq: diff-background}
     [\hat{\alpha} , -] = \theta_i \, [Z^i, -] = \theta_i V_{z^i},
\end{equation}
where $V_{z^i}$ is the Hamiltonian vector field with respect to the Poisson bivector $\vartheta$ of the $z_i = x^{2i-1} + i x^{2i}$ viewed as a coordinate on the worldvolume of the brane. Explicitly, we have
\begin{equation} \label{eq: Vi}
    V_{z^i} = i \, \vartheta^{2i-1, J} \, \partial_J - \vartheta^{2i, J} \, \partial_J =  \Theta^{ij} \, \partial_{z_j} + \Theta^i_j \, \partial_{\bar z_j} ,
\end{equation}
The vector fields $V^i$ span a distribution in the tangent space of the worldvolume of the brane which is always at least half dimensional, thereby identifying the worldvolume with a product $\RR^d \times \CC^k$.

For a brane background purely of type $(1,1)$, we diagonalize $\Theta^{(1,1)}$ so that the differential~\eqref{eq: diff-background} gives 
\begin{equation}
    [\hat{\alpha} , -] = \sum_{i = 1}^r \theta_i \partial _{\bar z^i}.
\end{equation}
Renaming $\theta_i$ to $\d \bar z_i$, we find that the space of fields is given by:
\begin{equation}
    (\Omega^{0,\bu}(\C^r) , \bar \partial) \otimes \C[\theta^{r+1}, \dots \theta^5] \otimes \mathfrak{gl}_k.
\end{equation}
Note that the Moyal product in this background trivializes in cohomology. In complex coordinates, the bivector entering~\eqref{eq: moyal} is
\begin{equation}
    \vartheta^{IJ} \, \partial_I \otimes \partial_J = - i \, \Theta^i_j \, \big( \partial_{z_i} \otimes \partial_{\bar z_j} - \partial_{\bar z_j} \otimes \partial_{z_i} \big) ,
\end{equation}
so that the Moyal product collapses to the ordinary multiplication of functions in $\bar \partial$-cohomology.

Let us now turn to branes that feature $\Theta^{(2,0)} \neq 0$. As discussed above, these branes are either Lagrangian or coisotropic depending on the form of $\Theta^{(1,1)}$. We start with the prototypical example arising in the $\SU(3)$-twist with commutators:
\begin{equation}
    [Z^1, Z^2] = i \vartheta \quad \text{and} \quad [Z^1, \bar Z_2] = [\bar Z_1, Z^2] = i \rho.
\end{equation}
The corresponding Hamiltonian vector fields take the form:
\begin{equation} \label{eq: V-su3}
    V_{z^1} = i \, (\vartheta \, \partial_{z_2} + \rho \, \partial_{\bar z_2}) , \qquad V_{z^2} = - i \, (\vartheta \, \partial_{z_1} + \rho \, \partial_{\bar z_1}) .
\end{equation}
For $\rho = \vartheta$, this simplifies to
\begin{equation}
    V_{z^1} = i \vartheta \, \partial_{x_3} , \qquad V_{z^2} = - i \vartheta \, \partial_{x_1}.
\end{equation}
Renaming $\d x_3 = i \vartheta \, \theta_1$ and $\d x_1 = - i \vartheta \, \theta_2$, we thus identify the BV complex in this twist as
\begin{equation}
    (\Omega^\bu(\R^2) , \d_{dR}) \otimes \C[\theta_3, \theta_4, \theta_5] \otimes \mathfrak{gl}_k .
\end{equation}
Lagrangian branes built from several such pairs behave in the same way; for instance the D3 brane in the $\mathrm{Sp}(4)_A$-twist leads to $(\Omega^\bu(\R^4), \d_{dR}) \otimes \C[\theta_5] \otimes \mathfrak{gl}_k$. The Moyal product is trivialized by the presence of the de Rham differential which only allows for constants in cohomology.

For $\rho \neq \pm \vartheta$, the worldvolume of the brane is four-dimensional along $\R^4_{x_1, \dots, x_4}$. The vector fields~\eqref{eq: V-su3} are complex and the distribution they span is half-dimensional so that they define a complex structure $I$ identifying $\RR^4 \cong \CC^2_I$. Note that this complex structure is always misaligned with the reference complex structure defined by the maximal isotropic $L$ since $\vartheta \neq 0$. With respect to $I$, the vector fields $V_{z^i}$ now define the $\bar \partial$-operator so that we can identify the space of fields as
\begin{equation}
    (\Omega^{0, \bu}(\C^2_I) , \bar \partial) \otimes \C[\theta_3, \theta_4, \theta_5] \otimes \mathfrak{gl}_k .
\end{equation}
In contrast to the previous two cases, the Moyal product here is non-trivial in cohomology. Let $w_a, \bar w_a$ denote complex coordinates adapted to the complex structure $I$ given by $w_a = \rho z_a - \vartheta \bar z_a$. Then the bivector in the definition of the Moyal product contains a term
\begin{equation}
    \vartheta^{IJ} \partial_J \sim \vartheta (\vartheta^2 - \rho^2) (\partial_{w_1} \otimes \partial_{w_2} - \partial_{w_2} \otimes \partial_{w_1})
\end{equation}
which acts non-trivial in cohomology. Explicitly, we find non-commutativity in these coordinates parametrized by
\begin{equation}
    [w_1, w_2]_{\star_\vartheta} = i \vartheta \, (\vartheta^2 - \rho^2),
\end{equation}
so that worldvolume of the coisotropic brane is a non-commutative $\C^2_I$.

The general picture we arrive at is thus as follows: Let $D$ be the distribution spanned by the vector fields $V_{z^i}$. Then the real part $D \cap \bar D$ defines topological directions along which the differential acts as the de Rham differential. On a complement, $D$ defines a complex structure which coincides with the original complex structure defined by $L$ if the background is purely of type $(1,1)$ in these directions and differs if it has a $(2,0)$ component.

In summary, twists of the IKKT model in block diagonal backgrounds are thus constructed as products,
\begin{equation}
    \Omega^{0,\bu}(\CC^r) \otimes \Omega^\bu(\R^{2s}) \otimes \Omega^{0,\bu}(\C^t_I)\otimes \C[\theta^{r+2s+t+1}, \dots, \theta^5] \otimes \mathfrak{gl}_k,
\end{equation}
where $r$ is the number of directions with a background purely of type $(1,1)$, the dimension of $D \cap \bar D$ is $2s$ and $t$ is the number of non-commutative directions with a complex structure misaligned to that defined by $L$. Finally, for commuting directions, we retain an odd variable $\theta^i$ and $k$ is additional data corresponding to the number of branes in the stack that the background describes.

For a pure dynamical supercharge, backgrounds that are not block diagonal lead to the same worldvolume theories. In general, $D$ and $\bar D$ span the complexified tangent space of the worldvolume, which therefore splits into the de Rham directions $D \cap \bar D$ and a complement with a complex structure. After a linear change of coordinates on the worldvolume, the differential takes the form above, and the product in cohomology is the Moyal product, which is non-trivial along $t = \rank \Theta^{(2,0)} - 2s$ of the complex directions and trivial along the remaining $r$. Since $\rank \Theta^{(2,0)}$ is fixed by the twisting supercharge, the possible values of $(r, s, t)$ are exactly those realized by block-diagonal backgrounds; off-diagonal terms only tilt the worldvolume relative to the topological and holomorphic directions.

This background expansion is the twisted counterpart of the emergence of noncommutative Yang–Mills theory on D-brane backgrounds in the IIB matrix model \cite{Aoki:1998bq, Aoki:1999vr}. The cohomological reduction here identifies the surviving BV fields and differential of the emergent worldvolume theory in the twist.

In fact, the same classical fluctuation calculation extends to a graded multiplicity space $V=\mathbb C^{k_+|k_-}$; this scenario arises commonly in twisted holography (e.g. \cite{Costello:2018zrm, Gaiotto:2024dwr}) when one studies both branes and ghost branes in the bulk \cite{Okuda:2006fb}. The background then has the form
$\alpha_0\otimes 1_V$, and
\[
 [\alpha_0\otimes 1_V,a\otimes M]
   =[\alpha_0,a]\otimes M .
\]
Therefore, the fields become valued in $\mathfrak{gl}(k_+|k_-)$, with the total
grading and the supertrace replacing the ordinary trace, but the complex and products are otherwise unchanged. The boundary state of the brane stack is instead weighted by $\operatorname{Str}(1_V)=k_+-k_-$. In particular, when $k_+ = k_-$ the closed source of this stack vanishes and the twisted branes do not backreact on BCOV theory \cite{CLBCOV1}.

\subsection{The holomorphic non-isotropic twist in component fields}
In the following, let us discuss some of the worldvolume theories in more detail and show how they arise from the component field description of the minimally twisted IKKT model. We start with the holomorphic non-isotropic twist and then move to the worldvolume theory on a Lagrangian brane along the A-model directions in~\S\ref{sec: a-model-comp}.

We move to a holomorphic non-isotropic twisting background by setting $\edyn = 1$ and $\ekin = 1$ together with a brane background satisfying
\begin{equation}
    [Z^a, \bar{Z}_a]= 2 .
\end{equation}
Here, we take the minimal choice setting  $[Z^1, \bar{Z}_1] = 2$ which corresponds to the stack of D-instantons forming a D1 brane. As discussed above, the twisted theory in this background is described by the BV complex,
\begin{equation} 
    \Omega^{0,\bu}(\C) \otimes \C[\theta^{2}, \dots, \theta^{5}] \otimes \mathfrak{gl}_k.
\end{equation}
Recall that this is the dimensional reduction of holomorphic Chern--Simons theory on $\C^5$, with the odd generators packaging the transverse directions to the brane, and therefore coincides with the worldvolume theory on a stack of $k$ B-branes placed along $\CC \subset \CC^5$.

Let us now derive this result from the component field description. For that purpose, let's fix a D1 brane background with $[\hat{Z}^1, \hat{\bar{Z}}_1] = 2$ and expand the holomorphic isotropic twist by replacing
\begin{equation}
    Z^1 \to \hat{Z}^1 +  Z^1.
\end{equation}
The adjoint action of the background $[\hat{Z}^1, -]$ under our identification becomes $\partial_{\bar z}$ on coefficient functions. We denote indices in the directions with no expectation values as $a=2,\dots ,5$. All other fields decompose accordingly, for example in the case of the fermion $\psi^{(2)}$ we have
\begin{equation}
    \psi^{(2) ij} \mapsto (\psi^{ab} , \psi^{1a} =: \psi^a).
\end{equation}
Expanding the BV action to quadratic order, we find the following linear transformations for the BV operator
\begin{equation}
    \begin{alignedat}{2}
    Q_{BV} c           &= 0 &\qquad\quad Q_{BV} Z^1 &= \bar\partial c \\
    Q_{BV} Z^a         &= 0 &\qquad\quad Q_{BV} \psi^a        &= \bar\partial Z^a \\
    Q_{BV} \psi^{ab}   &= 0 &\qquad\quad Q_{BV} \psi^{*ab}    &= \bar\partial \psi^{ab} \\
    Q_{BV} \psi^{*abc} &= 0 &\qquad\quad Q_{BV} Z^*_d         &= \varepsilon_{dabc} \, \bar\partial \psi^{*abc} \\
    Q_{BV} Z^*_1       &= 0 &\qquad\quad Q_{BV} c^*           &= \bar\partial Z^*_1
\end{alignedat}
\end{equation}

This identifies the space of BV fields as $\Omega^{0,\bu}(\CC) \otimes \C[\theta^2, \dots, \theta^5]$ explicitly along the following table.
\begin{table}[h]
\centering
\renewcommand{\arraystretch}{1.5}
\begin{tabular}{|c|c|c|c|c|c|}
\hline
 & $1$ & $(\theta^a)^1$ & $(\theta^a)^2$ & $(\theta^a)^3$ & $(\theta^a)^4$\\
\hline
1 & $c$ & $Z^a$ & $\psi^{ab}$ & $\psi^{*abc}$ & $Z_1^*$ \\
\hline
$ \d \bar{z}$ & $Z$ & $\psi^a$ & $\psi^{*ab}$ & $Z^*_a$ & $c^*$\\
\hline
\end{tabular}
\caption{Fields of the holomorphic non-isotropic twist. We denote $\psi^{* ab}:= \psi^{*1ab}$. We suppress the common $\mathfrak{gl}_k$ factor of all fields. The columns $(\theta^a)^p$ denote terms of transverse degree $p$, and hence products of distinct odd generators.}
\end{table}

The holomorphic non-isotropic twist of the IKKT model thus describes a situation in which the stack of $N\rightarrow \infty$ D-instantons dissolves to a stack of $k$ B-branes placed along a complex submanifold $\C^r \subset \C^5$. The rank $k$ of the B-brane worldvolume gauge algebra may remain finite, but it is interesting to study it in various limits. Taking a separate limit $k \rightarrow \infty$, gives the stable single-trace sector of the emergent B-brane theory on $\mathbb C^r\subset\mathbb C^5$. Applying the same open-closed map to this emergent B-brane stack constructs BCOV theory in the bulk as the closed string field theory\footnote{Similarly, the open-closed-open triality constructions in twisted holography \cite{Mazenc:2026yrm} involve distinct brane sectors probing the same backreacted closed string background.}. This also clarifies the relation between the two distinct orbits of holomorphic twists in the type IIB nilpotence variety~\cite{iib-nv}. The holomorphic isotropic twist in the bulk is dual to the minimal twist of IKKT which describes the theory on a stack of D-instantons in a background where only the dynamical supercharge takes a vev. In the holomorphic non-isotropic twist, these D-instantons have formed a higher-dimensional B-brane; in both cases the associated closed string field theories are BCOV theories.

\subsection{Lagrangian branes in component fields} \label{sec: a-model-comp}
Let us now derive the worldvolume theories on Lagrangian A-brane backgrounds from the minimally twisted IKKT model. Here, we restrict ourselves to the product of a Lagrangian A-brane with a point brane along the B-model directions. For the $\SU(3)$-twist featuring two non-commutative directions, we have four topological A-model directions and the brane is of the form $\RR^2 \times \{0\} \subset \RR^4 \times \C^3$. For the $\mathrm{Sp}(4)_A$-twist with four non-commutative directions, we find $\R^4 \times \{0\} \subset \R^8 \times \CC$.

We leave the component field derivation of the additional coisotropic branes in the $\SU(3)$ and $\mathrm{Sp}(4)_A$ twists and a more detailed comparison of their open string complexes with those of \cite{Kapustin:2005uy} to future work.

\subsubsection{The $\SU(3)$-twist}
Recall that this twist is described by $\edyn = 1$ and $\ekin = -i e^\vee_{12}$ together with
\begin{equation}
    [\hat{Z}^1, \hat{Z}^2] = i,
\end{equation}
where we have set the parameter $\vartheta=1$ for convenience. In terms of BCOV theory on $\CC^5$, this corresponds to a deformation by the bivector
\begin{equation}
    \partial_{z_1} \wedge \partial_{z_2} \in \mathrm{PV}^{(2,0)}(\C^5).
\end{equation}

We denote the two non-commutative directions along the brane by $\mu,\nu=1,2$ and the three commutative transverse directions by $a,b=3,\dots, 5$. The space of BV fields is
\begin{equation}\label{eq: bv-fields-su3}
	(\Omega^\bu(\RR^2) , \d_{dR} ) \otimes \CC[\theta_3, \theta_4, \theta_5] \otimes \mathfrak{gl}_k .
\end{equation}
This is the worldvolume theory on a stack of $k$ D1 branes placed along $\RR^2 \times \{ 0 \} \subset \RR^4 \times \C^3$ in a mixed A/B model.

Explicitly, this can be again seen by expanding the component fields by 
\begin{equation}
    Z^\mu \to \hat{Z}^\mu + Z^\mu.
\end{equation}
At the linearized level, we find the following for the BV operator in this background.
\begin{equation}
	\begin{alignedat}{3}
		Q_{BV} c &= 0 &\qquad Q_{BV} Z^\mu &= \partial^\mu c &\qquad Q_{BV} \psi^{12} &= \varepsilon_{\mu\nu} \, \partial^\mu Z^\nu \\
		Q_{BV} Z^a &= 0 & Q_{BV} \psi^{\mu a} &= \partial^\mu Z^a & Q_{BV} \psi^{*12a} &= \varepsilon_{\mu\nu} \, \partial^\mu \psi^{\nu a} \\
		Q_{BV} \psi^{ab} &= 0 & Q_{BV} \psi^{*\mu ab} &= \partial^\mu \psi^{ab} & Q_{BV} Z^*_c &= \tfrac12 \varepsilon_{cab} \varepsilon_{\mu\nu} \, \partial^\mu \psi^{*\nu ab} \\
		Q_{BV} \psi^{*345} &= 0 & Q_{BV} Z^*_\mu &= \varepsilon_{\mu\nu} \, \partial^\nu \psi^{*345} & Q_{BV} c^* &= \partial^\mu Z^*_\mu
	\end{alignedat}
\end{equation}
Each row in this equation represents a copy of the de Rham complex. For example, the first row is the familiar sequence from a scalar to a one-form and then to its curl. The component fields are identified with~\eqref{eq: bv-fields-su3} as laid out in the following table.
\begin{table}[h]
	\centering
	\renewcommand{\arraystretch}{1.5}
	\begin{tabular}{|c|c|c|c|c|}
		\hline
		& $1$ & $(\theta_a)^1$ & $(\theta_a)^2$ & $(\theta_a)^3$ \\
		\hline
		$\Omega^0$ & $c$ & $Z^a$ & $\psi^{ab}$ & $\psi^{*345}$ \\
		\hline
		$\Omega^1$ & $Z^\mu$ & $\psi^{\mu a}$ & $\psi^{*\mu ab}$ & $Z^*_\mu$ \\
		\hline
		$\Omega^2$ & $\psi^{12}$ & $\psi^{*12a}$ & $Z^*_a$ & $c^*$ \\
		\hline
	\end{tabular}
	\caption{Fields of the  $\SU(3)$ twist.}
	\label{tab: fields-su3}
\end{table}

\subsubsection{The $\mathrm{Sp}(4)_A$-twist}
Let now $\edyn = 1$ and $\ekin = -i (e^\vee_{12} + e^\vee_{34})$ together with the non-commutative background
\begin{equation}
    [Z^1, Z^2]  = [Z^3, Z^4] = i.
\end{equation}
In terms of BCOV theory on $\CC^5$, this corresponds to a deformation by the holomorphic bivector
\begin{equation}
    \partial_{z_1} \wedge \partial_{z_2} + \partial_{z_3} \wedge \partial_{z_4} \in \mathrm{PV}^{(2,0)}(\CC^5).
\end{equation}

The space of BV fields for this twist is
\begin{equation}
    (\Omega^\bu(\RR^4) , \d_{dR}) \otimes \CC[\theta^5] \otimes \mathfrak{gl}_k.
\end{equation}
This is the worldvolume theory of a stack of $k$ D3 branes placed along $\R^4 \times \{0\} \subset \RR^8 \times \CC$ in a mixed A/B model.

In component fields, we denote the non-commutative directions by $\mu, \nu, \dots = 1, \dots, 4$, expand as $\hat{Z}^\mu + Z^\mu$, and write $\partial^\mu = [\hat{Z}^\mu, -]$. Expanding the BV action, we find that the linearized BV operator acts schematically as follows:
\begin{equation} \label{eq: bv-sp4}
	\begin{tikzcd}[row sep=0.6em, column sep=1.9em]
		c \arrow[r, "\partial"] & Z^\mu \arrow[r, "\partial"] & \psi^{\mu\nu} \arrow[r, "\partial"] & \psi^{*\mu\nu\rho} \arrow[r, "\partial"] & Z^*_5 \\
		Z^5 \arrow[r, "\partial"] & \psi^{\mu 5} \arrow[r, "\partial"] & \psi^{*\mu\nu 5} \arrow[r, "\partial"] & Z^*_\sigma \arrow[r, "\partial"] & c^*
	\end{tikzcd}
\end{equation}
Each line is a copy of the de Rham complex of the worldvolume $\RR^4$, and the fields are organised as in Table~\ref{tab: fields-sp4}.
\begin{table}[h]
	\centering
	\renewcommand{\arraystretch}{1.5}
	\begin{tabular}{|c|c|c|c|c|c|}
		\hline
		& $\Omega^0$ & $\Omega^1$ & $\Omega^2$ & $\Omega^3$ & $\Omega^4$ \\
		\hline
		$1$ & $c$ & $Z^\mu$ & $\psi^{\mu\nu}$ & $\psi^{*\mu\nu\rho}$ & $Z^*_5$ \\
		\hline
		$\theta_5$ & $Z^5$ & $\psi^{\mu 5}$ & $\psi^{*\mu\nu 5}$ & $Z^*_\mu$ & $c^*$ \\
		\hline
	\end{tabular}
	\caption{Fields of the $\mathrm{Sp}(4)_A$ twist.}
	\label{tab: fields-sp4}
\end{table}

Explicitly, we can write $\alpha = A + \theta^5 B$ using our remaining odd generator to package the fields into two BV superfields, each of which is valued in the de Rham complex of $\mathbb{R}^4$. This is precisely the organization used in the BV description of four-dimensional BF theory:
\begin{equation}
    S_{BV} \propto \int_{\mathbb{R}^4}\textrm{tr}(B(dA + {1 \over 2}[A, A])).
\end{equation}
The physical fields include the one-form connection $A$ and the adjoint two-form $B$; the superfields also contain ghosts and antifields.

\subsection{Linear superpotentials and the IKKT partition function} \label{sec: spin7}
In addition to the worldvolume theories arising as twists of the IKKT model discussed in the previous sections, there are three twists which set the dynamical twisting supercharge to an impure spinor. Acting with $\Spin(10,\C)$ each impure spinor can be brought to the form $\edyn = 1 + e^\vee_{1234}$ so that these twists all are dual to linear superpotential deformations in the bulk by~\eqref{eq: res-bcov}.

The twists of this type are the $\Spin(7)$, $\mathrm{Sp}(4)_B$, and $\SU(4)$-twists, where the $\Spin(7)$ twist is the minimal twist of this class while the $\mathrm{Sp}(4)_B$- and $\SU(4)$-twists can be obtained as further deformations from it. As shown in~\cite{tw-bfss-ikkt}, setting the dynamical twisting supercharge to an impure spinor renders the BV complex acyclic so that these twists are trivial in perturbation theory. This corresponds to the bulk statement that turning on a linear superpotential in BCOV theory leaves no local degrees of freedom.

Even though the perturbative classical BV complexes of these twists are trivial, they feature interesting information beyond perturbation theory. Indeed, for the finite-$N$ $\Spin(7)$-twist, we can compute its (nontrivial) matrix integral. We consider the interacting $\SU(N)$ sector, with the decoupled $\U(1)$ center of mass modes removed. The partition function of the IKKT model
\begin{equation} \label{eq: part-func}
    Z_N = \frac{1}{\mathrm{Vol}(G)} \int \d^{10} X \, \d^{16} \psi \; e^{-S_0} , \qquad G = \SU(N)/\ZZ_N,
\end{equation}
was computed in~\cite{MNS} using supersymmetric localization. The result is
\begin{equation}\label{eq:divisorsum}
    Z_N = \sum_{d|N} \frac{1}{d^2}
\end{equation}
which is the D-instanton measure that Green and Gutperle extracted from the $R^4$ coupling of type IIB string theory~\cite{Green:1997tn}. The computation in~\cite{MNS} introduces an auxiliary field $\Vec{H}$ so that the square-zero symmetry acts on the physical fields of the IKKT model. We now show that an ``on-shell version'' of this computation naturally arises as a BV integral in a version of the $\Spin(7)$-twist deformed by an $\Omega$-background. More specifically, we will find that the quadratic antifield terms encode the closure implemented by the additional auxiliary fields in~\cite{MNS}. We will also identify the canonical transformation that places the matrix integral directly into its localized form. 

In the BV formalism, the partition function is represented as an integral over a Lagrangian\footnote{A Lagrangian in this context is a subspace of the extended field space on which the BV symplectic form vanishes.} $\cL \subset \cE_{BV}$ in the space of BV fields~\cite{BV1, Schwarz-BV},
\begin{equation}
    Z_N = \int_{\cL_0 \subset \cE_{BV}} e^{-S_{BV}},
\end{equation}
and~\eqref{eq: part-func} arises by choosing the trivial Lagrangian $\cL_0 = \{ (\Phi, \Phi^*) \in \cE_{BV} | \Phi^* = 0 \}$ defined by setting all antifields to zero. To compute this integral we perform three steps:
\begin{enumerate}
    \item We place the theory in the twisting background for the $\Spin(7)$-twisting supercharge, i.e. we set $\edyn = Q = 1 + e^\vee_{1234}$. Note that this does not change the integral, since all terms in $S_{\mathrm{susy}}$ are at least linear in the antifields and therefore vanish when restricted to $\cL_0$.
    \item We perform a canonical transformation on the space of BV fields so that the action simplifies.
    \item We deform the twisting supercharge $Q$ to an $\Omega$-background which changes the BV action again but keeps $Z_N$ constant. In this background, we can deform the Lagrangian $\cL_0$ to a family $\cL_t$ and the integral localizes in the limit $t \rightarrow \infty$. 
\end{enumerate}
Let us decompose the fields of the IKKT model under $\Spin(7)$ viewed as the stabilizer of an impure spinor in $S_+$ (see~\cite{tw-bfss-ikkt} for details). This can be done by first restricting to $\Spin(8) \times \Spin(2)$ along the standard embedding and then to $\Spin(7)$ along the embedding that preserves a positive chirality spinor. This gives the following inclusions of subgroups.
\begin{equation}
    \Spin(7) \times \U(1) \subset \Spin(8) \times \Spin(2) \subset \Spin(10).
\end{equation}
The vector and spinor representations of $\Spin(10)$ then decompose as:
\begin{equation}
    \begin{split}
        V &\longrightarrow \textbf{8}^0 \oplus \textbf{1}^{+2} \oplus \textbf{1}^{-2} \\
        S_+ &\longrightarrow \textbf{1}^{+1} \oplus \textbf{7}^{+1} \oplus \textbf{8}^{-1}.
    \end{split}
\end{equation}
Accordingly, we write for the field content:
\begin{equation}
    X^I \mapsto (X^a , Z_{-2} , \bar{Z}_{+2}) \qquad \psi^\alpha \mapsto (\psi_+, \psi_+^i, \psi_-^a),
\end{equation}
where $a$ is an index for the eight-dimensional spinor representation and $i$ for the seven-dimensional vector representation of $\Spin(7)$. The subscripts denote $U(1)$-charges of the fields. Note that the antifields carry opposite $U(1)$-charges, so that for instance $X^{*I}$ decomposes as $(X^{*a}, Z^*_{+2}, \bar{Z}^*_{-2})$

As usual, instead of integrating over the ghost field, we divide by the volume of the gauge group and eliminate $c$ and $c^*$ from the BV complex. Note that the twisting supercharge no longer squares to zero on the reduced complex. Instead we have
\begin{equation}
    (\delta_Q)^2 = [Z_{-2},- ] = \ad_{Z_{-2}},
\end{equation}
squaring to the adjoint action of $Z_{-2}$. This reflects the fact that the supersymmetry algebra only closes up to gauge transformations. In the following, the BV space of fields will therefore only consist of $(X, \psi)$ as well as their antifields.

We denote the BV action in the twisting background set by $\edyn = Q$ by $S_{BV}^Q$. Recall that it decomposes
\begin{equation}
    S_{BV}^Q = S_0 + S_{\mathrm{susy}}^{\mathrm{dyn}}|_{\edyn = Q}
\end{equation}
into the classical action $S_0$ and the contribution from the dynamical SUSY ghost set to the value of the twisting supercharge. Decomposing the action in terms of the $\Spin(7)$-equivariant fields gives for the classical action:
\begin{equation} \label{eq: S0-spin7}
\begin{aligned}
    S_0 = \tr \Big( & \tfrac{1}{128} [Z_{-2}, \bar{Z}_{+2}]^2 + \tfrac18 [Z_{-2}, X^a][X_a, \bar{Z}_{+2}] - \tfrac12 \cE^i \cE_i \\
    & - \tfrac{1}{128} \psi_+ [Z_{-2}, \psi_+] - \tfrac18 \psi^a_- [\psi_{a-}, \bar{Z}_{+2}] - \tfrac18 \psi^a_- [X_a, \psi_+] \\
    & + 2 \pi^i{}_{ab} \, \psi^i_+ [\psi^a_-, X^b] - \tfrac12 \psi^i_+ [Z_{-2}, \psi_{i+}] \Big) .
\end{aligned}
\end{equation}
Here $\pi^i_{ab}$ are the components of the Spin(7)-equivariant projection $\pi: \wedge^2 \textbf{8} \longrightarrow \textbf{7}$. For later reference, we define $\cE^i(X) = \pi^i_{ab} [X^a, X^b]$. Further, we find:
\begin{equation} \label{eq: Ssusy-spin7}
    S^Q_{\mathrm{susy}} = \tr \Big( \bar{Z}^*_{-2} \psi_+ + \psi^*_- [Z_{-2}, \bar{Z}_{+2}] + X^*_a \psi^a_- + \psi^{*a}_+ [Z_{-2}, X_a] + \psi^{*i}_- \cE_i \Big) - \tfrac12 \tr \big( \psi^{*i}_- \psi^{*}_{i-} \big) .
\end{equation}
In~\cite{MNS}, an auxiliary field $H \in \textbf{7}^0$ is introduced that satisfies $H^i = \cE^i - \psi^{* i}$ together with the supersymmetry transformation $\delta_Q \psi^i_+ = H^i$, $\delta_Q H^i = [Z_{-2}, \psi_+^i]$. Introducing the auxiliary field gives an off-shell completion, while eliminating it recovers the BV formulation with the quadratic antifield term.

The action of the twisting supercharge is encoded in $S_{\mathrm{susy}}^Q$. One finds the following transformations on the fields (which equivalently also arise as restrictions of~\eqref{eq: dyn-susy trafos} to $\edyn = Q$):
\begin{equation} \label{eq: spin7-trafos}
\begin{array}{l@{\qquad\qquad}l}
    \delta Z_{-2} = 0 & \delta Z^*_{+2} = [\bar{Z}_{+2}, \psi^*_-] + [X_a, \psi^{*a}_+] \\[6pt]
    \delta \bar{Z}_{+2} = \psi_+ & \delta \bar{Z}^*_{-2} = - [Z_{-2}, \psi^*_-] \\[6pt]
    \delta X^a = \psi^a_- & \delta X^*_a = - [Z_{-2}, \psi^*_{a+}] + 2 \pi^i_{ab} [X^b, \psi^*_{i-}] \\[6pt]
    \delta \psi_+ = [Z_{-2}, \bar{Z}_{+2}] & \delta \psi^*_- = - \bar{Z}^*_{-2} \\[6pt]
    \delta \psi^i_+ = \cE^i - \psi^{*i}_- & \delta \psi^*_{a+} = - X^*_a \\[6pt]
    \delta \psi^a_- = [Z_{-2}, X^a] & \delta \psi^*_{i-} = 0
\end{array}
\end{equation}
These are the on-shell versions of the square-zero symmetry used in~\cite{MNS}. With the gauge sector $c$ removed by hand, notice that $\delta^2 X^a = [Z_{-2}, X^a]$, a gauge transformation generated by $Z_{-2}$. It is therefore only nilpotent on gauge-invariant quantities. 

We now define a new set of coordinates on the space of BV fields to simplify the BV action. Generally, let us write $(\Phi, \Phi^*)$ for the collective coordinates of fields and antifields on $\cE_{BV}$. Given a function $W[\Phi]$ of ghost number $-1$ the change of coordinates
\begin{equation}
    (\Phi, \Phi^*) \mapsto (\Phi, \Tilde{\Phi}^* = \Phi^* + (-1)^{|\Phi|+1} \frac{\partial W}{\partial \Phi})
\end{equation}
is a canonical transformation on the space of BV fields, where $|\Phi|$ denotes the Grassmann parity of $\Phi$. Recall that the choice of twisting supercharge changes the ghost degree to $d_{gh} - d_{U(1)}$ which makes
\begin{equation} \label{eq: W-primitive}
    W = \tr \Big( \tfrac{1}{128} \psi_+ [Z_{-2}, \bar{Z}_{+2}] + \tfrac18 \psi^a_- [X_a, \bar{Z}_{+2}] - \psi^i_+ \cE_i \Big),
\end{equation}
a valid generator of such a canonical transformation. Expressing the BV action in these adapted coordinates, we find
\begin{equation} \label{eq: SQ-spin7}
\begin{aligned}
    \Tilde{S}(\Phi, \tilde\Phi^*) &= \tr \Big( \tilde{\bar{Z}}^*_{-2} \psi_+ + \tilde\psi^*_- [Z_{-2}, \bar{Z}_{+2}] + \tilde X^*_a \psi^a_- + \tilde\psi^{*a}_+ [Z_{-2}, X_a] \Big) \\
    & \quad - \tfrac12 \tr \big( \tilde\psi^{*i}_- \tilde\psi^{*}_{i-} \big) - \tfrac12 \tr \big( \psi^i_+ [Z_{-2}, \psi_{i+}] \big) .
\end{aligned}
\end{equation}
The canonical transformation has absorbed the bosonic potential and most Yukawa couplings into antifield-dependent terms, resulting in a simpler action. In particular, $\tilde\psi^{*i}_- = \psi^{*i}_- - \cE^i = - H^i$, so that the transformed antifield takes the role of the auxiliary field of~\cite{MNS}. This is the BV version of the cohomological reformulation used in localization. The original Lagrangian $\cL_0$ is expressed in new coordinates as $\Tilde{\Phi}^* = (-1)^{|\Phi|+1} \partial W/\partial \Phi$ so that
\begin{equation}
    Z_N = \int_{\tilde\Phi^* = (-1)^{|\Phi|+1} \partial W / \partial \Phi} e^{-\tilde S} .
\end{equation}
To perform the integral, we deform the twisting supercharge $Q$ to an $\Omega$-background. This then allows us to deform the Lagrangian $\cL_0$ to a family $\cL_t$ where the integral localizes once we take $t \rightarrow \infty$.

Recall that an $\Omega$-background on the supergravity side is a background where the local supersymmetry ghosts take a non-vanishing value that square to a rotation generating a circle action. Here, we want the rotation to be along the eight directions $X^a$ that transform under $\Spin(7)$. Thus, we are looking for a deformation of the twisting supercharge $Q$ to $Q_A$ so that
\begin{equation}
    (\delta_{Q_A})^2 = \textrm{ad}_{Z_{-2}} + A \qquad \text{for} \quad A \in \mathfrak{spin}(7)  
\end{equation}
holds on the space of IKKT fields. On gauge-invariant quantities the gauge transformation term drops out, leaving only the rotational action. This deformation can be achieved by the following modification of the BV action. Looking at~\eqref{eq: spin7-trafos}, we see that under the linearized transformations, the BV fields form pairs $(\Phi, \delta_Q \Phi)$ as follows:
\begin{equation}
    (\bar{Z}_{+2}, \psi_+) \qquad (X^a, \psi^a_-) \qquad (\psi^i_+ , -\psi^*_{-i}). 
\end{equation}
The first pair sits in the trivial representation of $\Spin(7)$ so that no modification is needed. For the other two, we add to the BV action
\begin{equation}
    \Tilde{S}^{A} = \Tilde{S}  + S_\Omega = \Tilde{S} + \tr(\tilde \psi^{*a}_+ \rho_{\textbf{8}}(A)_{ab} X^b ) - \frac{1}{2}  \tr (\psi_+^i \rho_{\textbf{7}}(A)_{ij} \psi_+^j ),
\end{equation}
where $\rho_{\textbf{7}}$ and $\rho_\textbf{8}$ are the representations by which $\Spin(7)$ acts on $\textbf{7}$ and $\textbf{8}$ respectively. The value of $Z_N$ does not change under the $\Omega$-deformation since $S_\Omega$ is of $U(1)$-weight $+2$ while the original integral is of weight zero. In the expansion of $e^{-S_{\Omega}}$, every nonconstant term therefore has vanishing expectation value in the undeformed, charge-invariant measure. The first term can be interpreted as assigning a value to the kinematical supersymmetry ghost that is linear in the eight bosonic matrices $X^a$, i.e. $\delta_{\Omega}\psi_-^a = \rho_8(A)^a_b X^b$. Note that restricting to $\cL_0$ given by $$\Tilde{\Phi}^* = (-1)^{|\Phi|+1} \partial W / \partial \Phi$$ gives
\begin{equation} \label{eq: S-Omega}
    S_\Omega|_{\cL_0} = - \tfrac18 \rho_{\textbf{8}}(A)_{ab} \tr \big( \bar Z_{+2} [X^a, X^b] \big) - \tfrac12 \tr \big( \psi^i_+ \rho_\textbf{7}(A)_{ij} \psi^j_+ \big) ,
\end{equation}
which is the deformation of the action found in~\cite{MNS}.

Choose a generic rotation $A$, which will have action on $\textbf{7}$ given by a one-dimensional fixed subspace. We will choose $n \in \textbf{7}$ a unit vector in that subspace. The subgroup fixing $n$ is $\textrm{Spin}(6) \simeq SU(4)$, under which we have the decomposition $\textbf{7} = \textbf{1}\oplus \textbf{6} = \langle n \rangle \oplus \textbf{6}$ and $\textbf{8}_{\mathbb{C}} = \textbf{4}\oplus \bar{\textbf{4}}$. According to this splitting, let us define:
\begin{equation} \label{eq: spin7-n}
    \psi^0_+ = n_i \psi^i_+ , \qquad \cE^0 = n_i \cE^i , \qquad J_{ab} = 2 n_i \pi^i_{ab} .
\end{equation}
Note that $J$ defines a complex structure on $\textbf{8}$ so that it decomposes into a copy of the fundamental and the antifundamental representation of $\mathrm{SU}(4)$.
We now deform $\cL_0$ to a family $\cL_t$ defined by
\begin{equation}
    \cL_t= \{ (\Phi , \Phi^*) | \Phi^* = (-1)^{|\Phi|+1} t\frac{\partial V}{\partial \Phi} \} = \{ (\Phi , \Tilde{\Phi}^*) | \Tilde{\Phi}^* = (-1)^{|\Phi|+1} \Big( \frac{\partial W}{\partial \Phi} +  t\frac{\partial V}{\partial \Phi} \Big) \},
\end{equation}
where
\begin{equation} \label{eq: spin7-V}
    V = n_i \tr \big( 2 i \pi^i_{ab} X^a \psi^b_- + \psi^i_+ \bar Z_{+2} \big) = \tr \big( i J_{ab} X^a \psi^b_- + \psi^0_+ \bar Z_{+2} \big).
\end{equation}
Notice that the two terms have ghost degrees $1$ and $-3$ respectively so that $V$ only preserves a total $\ZZ_2$-grading.\footnote{If one desires to restore the $\ZZ$-grading, we note that the equivariant parameters $A$ carry $U(1)$ charge $-2$. With $c(A)$ a scalar function scaling linearly in $A$, we can rescale the two terms by $c^{-1}$ and $c$ respectively thereby assigning twisted ghost degree $-1$ to $V$.}

We now restrict the $\Omega$-deformed action to $\cL_t$. The action separates into terms that are independent, linear and quadratic in  $t$. To take the limit $t \rightarrow \infty$, we rescale the fields by
\begin{equation} \label{eq: spin7-rescale}
    X^a \mapsto t^{-1/2} X^a , \qquad \psi^a_- \mapsto t^{-1/2} \psi^a_- , \qquad \bar Z_{+2} \mapsto t^{-1} \bar Z_{+2} , \qquad \psi_+ \mapsto t^{-1} \psi_+ .
\end{equation} 
This leaves the measure invariant since the contributions from bosons and fermions to the Jacobian cancel. After this field redefinition, the action separates into terms that are constant with respect to $t$ or scale as $1/t$ or $1/t^2$. Once we take the limit $t \rightarrow \infty$, only the constant part contributes, which is of the form
\begin{equation}
\begin{aligned}
    S_{t = \infty} =& \tr \big( i J_{ab} X^a ( [Z_{-2}, X^b] + \rho_\textbf{8}(A)^b_{\ c} X^c )\big)
    - \tfrac12 \tr \big( \psi^i_+ [Z_{-2}, \psi_{i+}] + \psi^i_+ \rho_\textbf{7}(A)_{ij} \psi^j_+ \big)  \\
    &+ \tr(i J_{ab} \psi^a_- \psi^b_- - \psi^0_+ \psi_+ - \tfrac12 \bar Z_{+2}^2 )
\end{aligned}
\end{equation}
For fixed $Z_{-2}$ these are Gaussian integrals; note that $\det(iJ_{ab}) = 1$ so that we obtain
\begin{equation} \label{eq: spin7-Z-int}
    Z_N \propto \frac{1}{\mathrm{Vol}(G)} \int \d Z_{-2} \, \frac{\mathrm{det}_{\textbf{6} \otimes \mathfrak{g}} (\ad_{Z_{-2}} + \rho_\textbf{6}(A))^{1/2}}{\det_{\textbf{8} \otimes \mathfrak{g}} (\ad_{Z_{-2}} + \rho_\textbf{8}(A) )^{1/2}} .
\end{equation}
This expression only depends on the eigenvalues of $A$. We can conjugate $A$ so that it lies in the maximal torus of $\mathrm{SU}(4)$ where it is parametrized by $(\epsilon_1, \dots , \epsilon_4)$ satisfying $\epsilon_1 + \dots + \epsilon_4 = 0$. This recovers the Gaussian reduction and determinant ratio used in \cite{MNS}. The remaining integral, with chosen contour prescription and measure, gives the result \eqref{eq:divisorsum}. The integration prescription must be discussed in addition to the classical BV data discussed in this section\footnote{A related distinction appears in polarized IKKT \cite{Komatsu:2024ydh}, which requires a further contour prescription to recover the undeformed partition function from its zero-mass limit.}.

It would be interesting to provide a geometric interpretation to this computation in terms of a non-perturbative computation in the holographically dual twist of supergravity and to extend toward the $\mathrm{Sp}(4)_B$- and $\SU(4)$-twists which arise by further deforming the $\Spin(7)$-twisting supercharge and introducing a brane background.

\printbibliography

\end{document}